\documentclass[twocolumn,aps,prd,superscriptaddress,nofootinbib]{revtex4-1}
\usepackage{graphicx}
\usepackage{amsmath}
\usepackage{amssymb}
\usepackage{wasysym}
\usepackage{bm}
\usepackage{slashed}
\usepackage{epsfig}
\usepackage{amsfonts}
\usepackage{epstopdf}
\usepackage{hyperref}
\usepackage[subfigure]{graphfig}
\usepackage{braket}
\usepackage{tikz}
\usepackage{hhline}
\usepackage{float}

\newcommand{\ben}{\begin{eqnarray}}
\newcommand{\een}{\end{eqnarray}}

\newcommand{\bef}{\begin{figure}[!htp]}
\newcommand{\eef}{\end{figure}}

\newcommand{\nn}{\nonumber}

\def\la{\label}
\newcommand{\bea}{\begin{eqnarray}}
\newcommand{\eea}{\end{eqnarray}}

\def\ba{\begin{linenomath*}\begin{equation}}
\def\ea{\end{equation}\end{linenomath*}}

\usepackage{soul}

\hypersetup{colorlinks, linkcolor = [rgb]{0,0.0,1.0}, citecolor = [rgb]{0,0.0,1.0}, urlcolor = [rgb]{0,0.0,1.0}}

\begin{document}

\null\hfill\begin{tabular}[t]{l@{}}
  {JLAB-THY-19-3038} 
\end{tabular}
\preprint{JLAB-THY-19-3038}

\title{Pion Valence Structure from Ioffe Time Pseudo-Distributions}

\newcommand*{\JLAB}{Thomas Jefferson National Accelerator Facility, Newport News, VA 23606, USA}\affiliation{\JLAB}
\newcommand*{\WM}{Physics Department, College of William and Mary, Williamsburg, Virginia 23187, USA}\affiliation{\WM}
\newcommand*{\CU}{Physics Department, Columbia University, New York City, New York 10027, USA}\affiliation{\CU}
\newcommand*{\ODU}{Physics Department, Old Dominion University, Norfolk, VA 23529, USA}\affiliation{\ODU}
\newcommand*{\Heidelberg}{Institute for Theoretical Physics, Heidelberg University, Philosophenweg 12, 69120 Heidelberg,Germany}\affiliation{\Heidelberg}
\newcommand*{\Marseille}{Aix Marseille Univ, Universit\'e de Toulon, CNRS, CPT, Marseille, France}\affiliation{\Marseille}

\author{B\'alint Jo\'o}\affiliation{\JLAB}
\author{Joseph Karpie}\affiliation{\WM}\affiliation{\CU}
\author{Kostas Orginos}\affiliation{\JLAB}\affiliation{\WM}
\author{Anatoly V. Radyushkin}\affiliation{\JLAB}\affiliation{\ODU}
\author{David G. Richards}\affiliation{\JLAB}
\author{Raza Sabbir Sufian}\affiliation{\JLAB}
\author{Savvas Zafeiropoulos}\affiliation{\Heidelberg},\affiliation{\Marseille}


\begin{abstract}
We present a calculation of the pion valence quark distribution extracted using the formalism of reduced Ioffe time pseudo-distributions or more commonly known as pseudo-PDFs. Our calculation is carried out on two different 2+1 flavor QCD ensembles using the isotropic-clover fermion action, with lattice dimensions $24^3\times 64$ and $32^3\times 96$ at  the lattice spacing of \mbox{$a=0.127$ fm}, and with the quark mass equivalent to a pion mass of $m_\pi \simeq 415$ MeV. We incorporate several combinations of smeared-point and smeared-smeared pion source-sink interpolation fields in obtaining the lattice QCD matrix elements using the summation method. After one-loop perturbative matching and combining the pseudo-distributions from these two ensembles, we extract the pion valence quark distribution using a phenomenological functional form motivated by the global fits of parton distribution functions. We also calculate the lowest four moments of the pion quark distribution through the ``OPE without OPE". We present a qualitative comparison between our lattice QCD extraction of the pion valence quark distribution with that obtained from global fits and previous lattice QCD calculations.  
\end{abstract}

\maketitle
\allowdisplaybreaks


\section{Introduction}	
\label{sec:intro}

The key element of most predictions involving hard inclusive reactions in high-energy physics is the factorization theorem~\cite{Collins:1989gx} of perturbative QCD. This factorization procedure separates the perturbatively calculable hard-scattering quark and gluon dynamics from the nonperturbative bound-state dynamics,  described  by the parton distribution functions (PDFs) of the relevant hadrons.

 For the nucleon, information about  the quark PDFs 
can be  obtained from the experimental data  of   deep inelastic scattering. 
 Numerous experiments have been performed, and the  functional form of the valence quark  PDFs is well understood in various global fits~\cite{Harland-Lang:2014zoa,Dulat:2015mca,Ball:2017nwa,Alekhin:2017kpj,Ethier:2017zbq}.
 
 On the other hand, the pion valence PDF has been extracted using the data from only a few pionic Drell-Yan experiments at CERN~\cite{Badier:1983mj,Betev:1985pf} and Fermilab~\cite{Conway:1989fs}. The valence PDF of the pion is of particular theoretical interest, as the pion is the lightest QCD bound state and the Goldstone mode associated with the spontaneous breaking of chiral symmetry. 
 
 The accurate form of the pion's valence quark distribution therefore provides a testing ground for both QCD and QCD-based approaches in understanding the structure of hadrons. The experimental data
 of Refs.~\cite{Badier:1983mj,Betev:1985pf,Conway:1989fs} have been analyzed in Refs.~\cite{Owens:1984zj,Aurenche:1989sx,Sutton:1991ay,Gluck:1991ey,Wijesooriya:2005ir,Aicher:2010cb,Barry:2018ort} to determine the pion valence distribution,  and compare, in particular, the  results of these analyses at large $x$ (fraction of the hadron's longitudinal momentum carried by the parton) with  the predictions  of  QCD-based  hard-gluon-exchange  models~\cite{Farrar:1979aw,Berger:1979du,Brodsky:1994kg}.  
 
 The large-$x$ behavior of the pion valence distribution has also been studied in different model \mbox{calculations~\cite{Shigetani:1993dx,Davidson:1994uv,Hecht:2000xa,Chen:2016sno,deTeramond:2018ecg,Ding:2019lwe,Lan:2019vui}}. However, despite having different fits to the experimental data and model calculations, it has not yet been settled whether the pion valence distribution near $x\to 1$  falls off as $(1-x)$ or follows the 
 $(1-x)^2$  behavior suggested
  in Refs.~\cite{Farrar:1979aw,Berger:1979du,Brodsky:1994kg}. 
  
   Thus, the slope of the pion valence quark distribution as $x\to 1$ may provide  important information about the 
  interplay of perturbative and nonperturbative aspects of 
   quark dynamics in the valence region. Because of its importance, understanding the large-$x$ behavior of the pion valence quark distribution is the goal of the approved experiment C12-15-006 at Jefferson Lab~\cite{Jlab}. 
   Intensive studies of the pion structure are also proposed for the future Electron-Ion  Collider~\cite{Aguilar:2019teb}.

   Another way of studying the behavior of the pion valence quark distribution may be provided 
  by lattice QCD calculations which can be complementary to the global fit analyses of the cross-section data 
  and can also serve as  a discriminator between different model predictions.  
  
  To achieve such a goal, we need  to go beyond the conventional moment calculations and  obtain $x$-dependent parton distributions.  Several new approaches that allow us to determine $x$-dependent
  parton distributions from lattice QCD have been proposed. These
  approaches include  the path-integral formulation of the deep-inelastic scattering hadronic tensor~\cite{Liu:1993cv}, the inversion method~\cite{Horsley:2012pz}, quasi-PDFs~\cite{Ji:2013dva}, good lattice cross sections~\cite{Ma:2014jla,Ma:2017pxb}, and reduced Ioffe time pseudo-distributions (or pseudo-PDFs)~\cite{Radyushkin:2017cyf,Radyushkin:2017sfi}.   
  An analogous coordinate-space method has been  earlier introduced for the
  calculation of light-cone distribution amplitudes~\cite{Braun:2007wv}.

  Although significant achievements in the lattice QCD implementations of these approaches have been made in recent years~\cite{Chambers:2017dov,Orginos:2017kos,Karpie:2017bzm,Alexandrou:2018pbm,Bali:2018spj,Bali:2019ecy,Lin:2018qky,Fan:2018dxu,Sufian:2019bol,Bali:2019dqc,Izubuchi:2019lyk,Liang:2019frk}, a proper understanding and controlling various sources of systematics in these calculations   still require further exploration and theoretical development. 
  The status of
    current lattice QCD calculations of the $x$-dependent hadronic
    structure can be found in the following review
    articles~\cite{Lin:2017snn,Monahan:2018euv,Cichy:2018mum}. Recently,
    an attempt to incorporate lattice QCD determination of PDFs
    together with experimental data to obtain the non-singlet quark
    distribution of the nucleon has been discussed
    in~\cite{Cichy:2019ebf}.

 Lattice calculations of the pion valence PDFs have been recently performed in 
Refs.~\cite{Chen:2018fwa,Sufian:2019bol,Izubuchi:2019lyk} using the quasi-PDF~\cite{Chen:2018fwa,Izubuchi:2019lyk}
and the good lattice cross sections~\cite{Sufian:2019bol}  approaches. In this  paper,   we present the first  lattice calculation of the pion valence PDF  using the
 approach based on  reduced Ioffe-time pseudo-distributions~\cite{Radyushkin:2017cyf}.  We discuss the  limitations  of our lattice QCD calculation  of the pion valence quark distribution and compare our results with 
 those in Refs.~\cite{Chen:2018fwa,Sufian:2019bol,Izubuchi:2019lyk} and also those obtained from 
  global fits as mentioned above.

The remainder of this article is organized as follows. In Section~\ref{method}, we briefly discuss the reduced Ioffe time pseudo-distributions approach and the necessity of the calculation in coordinate space.  In Section~\ref{numerical}, we present numerical details of the calculation of  hadronic matrix elements of the reduced Ioffe time pseudo-distribution to extract the pion valence quark distribution. We present the results of the extracted pion valence quark distribution in Section~\ref{results} and compare our result with different fits of the experimental data and other lattice calculations  in Section~\ref{comparison}. Finally, we summarize our results and outline the future directions of this method to obtain
the  pion valence quark distribution with controlled systematics.


\section{Basics of the Ioffe time pseudo-distributions approach} \la{method}

The unpolarized quark non-singlet PDF is defined as a Fourier transform of the 
 nonlocal matrix element
\bea \label{eq:matelem}
M^\alpha(p,z) \equiv \bra{p}\bar{\psi}(0)\gamma^\alpha\mathcal{W}(z;0) \psi(z) \ket{p}
\eea
taken on the light cone, e.g. for $z =(z_+=0,z_{-},0_\perp)$, with the momentum given by 
 $p =(p^+,p^-={m^2}/{2p^+},0_\perp)$ and $\alpha = +$,
 where we
  use the light-cone coordinates, $x_\pm = \frac{x_0 \pm x_3}{\sqrt 2}$.
The combination  $\nu=p\cdot z$ is called the Ioffe time~\cite{Ioffe:1969kf}, and  $\mathcal{W}(z;0)$ is the  gauge link in the fundamental representation. Its path goes along a straight-line $0\to z$. For general $z$, $p$ and $\alpha$, the Lorentz decomposition of this matrix element can be written as
\bea
M^\alpha(p,z) = 2p^\alpha \mathcal{M}(\nu,z^2)+2z^\alpha \mathcal{N}(\nu,z^2) \ .
\label{ITD}
\eea
When $z=z_-$ and $\alpha=+$, the second function $ \mathcal{N}(\nu,z^2)$ does not contribute,
i.e. the twist-2 PDF is solely determined by the first function. 
On the lattice, we need to take a spacelike $z$. Choosing  $z=z_3$ 
and $p = (E,0_\perp, p_3,)$ we can exclude the $ \mathcal{N}(\nu,z^2)$
function and deal with  
the function $\mathcal{M}(\nu,z^2)$ that  is called the Ioffe time pseudo-distribution (pseudo-ITD).  
The term ``pseudo"~\cite{Radyushkin:2017cyf} reflects the fact that one deals with the   matrix element
off the light-cone, i.e., for  nonzero $z^2$. 
Another  advantage  of taking the time component of $M^\alpha(p,z)$  on the lattice is that 
such a choice  also avoids the pitfall of renormalization constant mixing as described in Ref.~\cite{Constantinou:2017sej}.

 Choosing a spacelike separation $z$ brings in 
  a serious complication of additional link-related ultraviolet (UV) divergences~\cite{Polyakov:1980ca} 
  that are absent
  when $z$ is on the light cone. 
   Fortunately, these divergences are multiplicatively renormalizable~\cite{Ji:2017oey,Green:2017xeu,Izubuchi:2018srq},
   i.e. form an overall factor $Z(z^2/a^2)$, where $a$ is  a UV regulator, such as the lattice spacing.
    In the quasi-PDF approach, 
 such divergences  are usually  removed  using  various versions 
of the RI/MOM method~\cite{Alexandrou:2017huk,Chen:2017mzz}.

A different approach was proposed in~\cite{Radyushkin:2017cyf}. One considers the reduced pseudo-ITD
\bea \la{ratio}
\mathfrak{M}(\nu,z^2) = \frac{\mathcal{M}(\nu, z^2)}{\mathcal{M}(0,z^2)},
\eea
formed by the  ratio of $\mathcal{M}(\nu, z^2)$ to its rest-frame \mbox{$p_z=0$} value $\mathcal{M}(0, z^2)$.
Since the UV factor $Z(z^2/a^2)$ does not depend on $\nu$, it disappears from the ratio 
$\mathfrak{M}(\nu,z^2)$. As a result, the latter is UV-finite.
Moreover, it is a renormalization group invariant quantity. 
Also, taking the ratio \eqref{ratio}  removes not only the UV-divergences,
but also the part of the $z^2$-dependence
associated with them.

Beside UV divergences, there are other   sources of the $z^2$-dependence.
In particular,  the  function $\mathfrak{M}(\nu,z^2)$ contains higher-twist contributions 
related to the transverse-momentum distributions of quarks inside a hadron
and reflected in the $z^2$-dependence of $\mathcal{M}(\nu,z^2)$. 
For small $z^2$, they appear as   higher twist contributions 
that are polynomial in $\mathcal{O}(z^2 \Lambda^2_{\rm QCD})$.
Thus, one may expect that for small enough  $z^2$ they  may be neglected, and 
  $\mathcal{M}(\nu,z^2)$ may be related to the light-cone 
 Ioffe time distribution~\cite{Braun:1994jq} (ITD)  $Q(\nu, \mu^2)$.
 Such a relation is given by a factorization formula that involves 
 just logarithmic $\ln (z^2)$ dependence accompanied by 
  a perturbatively calculable kernel. 
  
  An alternative proposal of taking such a ratio similar to that in Eq.~\eqref{ratio} has been proposed in~\cite{Braun:2018brg}. In that article, the authors claim that a vacuum matrix element of the same type of operator could be used in the denominator, instead of the rest frame hadron matrix element.

  Furthermore, it is not unreasonable to suppose  that 
  the higher-twist $z^2$-dependence of $\mathcal{M}(\nu,z^2)$ 
  and $\mathcal{M}(0,z^2)$  is similar,  and 
  the higher-twist impact on the ratio $\mathfrak{M}(\nu,z^2)$
  is much weaker than $\mathcal{M}(\nu,z^2)$. In particular, if $\mathcal{M}(\nu,z^2)$
  factorizes like $\mathcal{M}(\nu,z^2) =\mathcal{M}(\nu) B(z^2)$,
  the ratio $\mathfrak{M}(\nu,z^2)$ has no \mbox{$z^2$-dependence. }
Unfortunately this idealized scenario is not true for QCD, but as was shown in~\cite{Karpie:2018zaz}, taking the ratio will always reduce the higher twist contribution, in the limit $\nu \to 0$. 
An actual calculation~\cite{Orginos:2017kos}, though performed in the quenched approximation, 
produced an almost $z^2$-independent result for  $\mathfrak{M}(\nu,z^2)$ in the region
$z>4a$ (i.e., in the region where one would expect higher twist  effects to be significant), 
and a logarithmic $\ln (z^2)$ dependence in the region $z \leq 4a$,
where it was perfectly described by the DGLAP (Dokshitzer-Gribov-Lipatov-Altarelli-Parisi)~\cite{Gribov:1972ri,Altarelli:1977zs,Dokshitzer:1977sg} evolution.

Summarizing,  the  choice of  $\mathfrak{M}(\nu,z^2)$ 
as the basic object for lattice studies of parton distribution functions satisfies  the following  criteria: 

\vspace{0.15cm}

\textcircled{\small{1}} Due to Lorentz invariance,  the pseudo-ITD  $\mathcal{M}(\nu,z^2)$ and, hence the reduced pseudo-ITD $\mathfrak{M}(\nu,z^2)$ 
depend only on the interval $z^2$   for a   spacelike separation $z$ between the quarks and the Ioffe time $\nu$.
Both  $z^2$ and $\nu$ are Lorentz-invariant.

\vspace{0.15cm}

\textcircled{\small{2}} For all values of spacelike $z^2$ and for any contributing Feynman diagram,
it can be shown~\cite{Radyushkin:2017cyf}  that  the Fourier transform of $\mathcal{M}(\nu,z^2)$ and, hence  $\mathfrak{M}(\nu,z^2)$)  
with respect to $\nu$  has  canonical support $-1\leq x \leq 1$ in the variable $x$
 interpreted as the standard momentum
  fraction.

\vspace{0.15cm}
\textcircled{\small{3}} The UV divergences associated with the gauge link are canceled in the 
reduced ITD,  and the latter is  a renormalization group invariant quantity.

\vspace{0.15cm}
\textcircled{\small{4}} The short-distance behavior due to $\ln (z_3^2M^2)$ terms
(where $M$ is an infrared regulator and $z=z_3$) 
present in  $\mathcal{M}(\nu,z_3^2)$ and  generating the perturbative evolution of parton densities 
is preserved in  $\mathfrak{M}(\nu,z_3^2)$.  In the $z_3^2\to 0$  limit, the reduced pseudo-ITD maps to the usual 
light-cone  PDF and obeys the  familiar   perturbative  DGLAP evolution with $1/z^2$ 
serving as an evolution parameter. 
\vspace{0.15cm}

To get the matching condition between  $\mathfrak{M}(\nu,z_3^2)$  
and the light-cone ITD $Q(\nu,\mu^2)$, one may  use the operator product expansion (OPE)  which is valid both 
for the numerator $\mathcal{M}(\nu,z_3^2)$ and the denominator $\mathcal{M}(0,z_3^2)$.
As discussed, the UV-singular factors present in these functions cancel
together with the $z_3^2$-dependence  associated with them. 
The remaining  $z_3^2$-dependence  corresponds to  the DGLAP logarithms $\ln (z_3^2M^2)$ and higher twist effects proportional to $z_3^2$.

In our particular case, we have an additional simplification  that, in the local limit,  the operator in Eq.(\ref{ITD}) 
is a conserved vector current. As a result,  the denominator  does not bring an extra $\ln (z_3^2M^2)$-dependence. 
Eventually,  $\mathfrak{M}(\nu,z_3^2)$  is matched  to the  $\overline{\rm MS}$ light-cone   ITD by
\bea \la{eq:factorizable}
\mathfrak{M}(\nu,z^2) = \int_0^1 du~ C(u,\mu^2z^2) Q(u\nu,\mu) + \mathcal{O}(z^2\Lambda_{\rm QCD}^2 ),\nn \\
\eea
where $Q(\nu,\mu)$ is the light-cone  ITD whose Fourier transform with respect to $\nu$ gives  the PDF $f(x,\mu)$ at a factorization scale $\mu$. The matching kernel $C(u,\mu^2z^2)$ of the reduced pseudo-ITD to the $\overline{\rm MS}$  ITD 
has  been  determined from  one-loop calculations~\cite{Radyushkin:2017lvu,Zhang:2018ggy,Radyushkin:2018cvn,Izubuchi:2018srq} 
\bea \la{evolution}
C(u,\mu^2z^2)&= &\delta(1-u) + \nn \\
&&\frac{\alpha_s C_F}{2\pi} \left[ \ln \left (z^2\mu^2 \frac{e^{2\gamma_E +1}}4 \right ) B(u) + L(u) \right]   \, ,\nn \\
\eea
where 
\bea
B(u) = \left[ \frac{1+u^2}{1-u}\right]_+
\eea
is the Altarelli-Parisi kernel~\cite{Altarelli:1977zs} and
\bea
L(u) = \left[ 4 \frac{\ln(1-u)}{1-u} - 2(1-u) \right]_+\,.
\eea
The inverse of this formula will be needed for converting the lattice results for  $\mathfrak{M}(\nu,z^2)$ to the $\overline{\mbox{MS}}$ ITD $Q(\nu,\mu)$ at a matching scale $\mu$. Without   loss of accuracy, this can be done 
by switching the reduced pseudo-ITD and the ITD and by changing the sign of $\alpha_s$.


%
\section{Numerical Methods} 
\label{numerical}

The pion reduced pseudo-ITD is calculated on two lattice QCD ensembles with different physical volumes. These configurations were generated by the JLab/W\&M collaboration~\cite{lattices} using 2+1 flavors of stout smeared clover Wilson fermions and a tree-level tadpole-improved Symanzik gauge action. The strange quark mass was set by requiring the ratio $\left(2M^2_{K^+}-M^2_{\pi^+}\right)/M_{\Omega^-}$ to assume its physical value. 

The fermion action includes one iteration of stout smearing with the weight for the staples given by \mbox{$\rho=0.125$}. This smearing procedure has the consequence that the tadpole corrected tree-level clover coefficient, $c_{\rm SW} = 1.2493$, is very close to the non-pertubative value determined a posteriori with the Schr\"odinger functional method~\cite{lattices}. The ensemble parameters are listed in Table~\ref{tab:lat}. The lattice spacing of these ensembles, \mbox{$a=0.127$ fm}, was determined using the Wilson flow scale $w_0$~\cite{Borsanyi:2012zs}.

\begin{table*}[t]
  \centering
  \begin{tabular}{|c|c|c|c|c|c|c|c|}
  \hline
    ID  & $a$ (fm)  &$m_\pi$ (MeV) &  $\beta$ & $am_l$ & $am_s$ & $L^3\times N_t$ & $N_{\rm cfg}$\\
    \hhline{|=|=|=|=|=|=|=|=|}
    $a127m415$ & $0.127(2)$ & 415(23) & 6.1 & -0.280 & -0.245 & $24^3\times 64$ & 2147 \\
    \hline
    $a127m415L$ & $0.127(2)$ & 415(23) & 6.1 & -0.280 & -0.245 & $32^3\times 96$ & 2560 \\
    \hhline{|=|=|=|=|=|=|=|=|}
  \end{tabular}
  \caption{The parameters for the JLab/W\&M collaboration ensembles 
  used in this work: lattice spacing, pion mass, $\beta$, light and strange quark mass, spatial and temporal size, and the numbers of configurations. The $a127m415L$ ensemble  contains 10 independent streams of 256 configurations each while the $a127m415$ ensemble contains a single stream. }
  \label{tab:lat}
\end{table*}

To ameliorate the contamination of excited states and improve the overlap of the interpolators onto boosted pions, we implement a combination of the Gaussian smearing~\cite{ Allton:1993wc} and momentum-smearing~\cite{Bali:2016lva} techniques. The pseudo-ITD matrix elements are calculated using the summation method. By now, this method is well known in the lattice community. For completeness, some key points of the method are highlighted below. We refer the readers to~\cite{Bouchard:2016heu,Orginos:2017kos,Chang:2018uxx,Joo:2019jct} for more details on the implementation of this method. 

The summation method is related to the Feynman-Hellmann theorem. One considers a theory where the action is modified by the operator of interest
\bea
S_\lambda (x) = S_{\rm QCD}(x) + \lambda \int d^4x O(x)\,.
\eea
By the Feynman-Hellmann theorem, a hadron matrix element of that operator can be found from a derivative of the energy of that hadron
\bea
\frac{d E_\lambda}{d\lambda} = \langle E_\lambda | \frac{d H_\lambda}{d\lambda} | E_\lambda\rangle\,.
\eea
As was shown in~\cite{Bouchard:2016heu}, the derivative of the effective mass can be shown to be a ratio of correlation functions. This method has an advantage over other methods based around the Feynman-Hellmann method, which require generation of specialized configurations which scan $\lambda$. The derivative of the effective mass at $\lambda=0$ can be calculated using standard gauge configurations with $\lambda=0$.

The two-point and three-point correlation functions for a fixed pion momentum $p$ and a fixed current insertion-time $t$ are written in terms of  the standard pion interpolation field $O_\Pi$
\bea
C_2(T) &=& \langle O_\Pi(T) \bar{O}_\Pi(0)\rangle \\
C_3(z,T) &=& \langle O_\Pi(T) \mathcal{O}_{\Gamma}(z)\bar{O}_\Pi(0)\rangle
\eea
where $T$ is the Euclidean time separation between the interpolating operators for the pion creation and annihilation operators, $\mathcal{O}_{\Gamma}(z)=\overline{\psi}(0)\Gamma\mathcal{W}(0;z)\psi(z)$, and we use $\Gamma=\gamma_4$.  For a fixed $z$, summing over the current insertion time $t$, the matrix element is estimated from the large Euclidean time limit of the effective matrix element
\bea
M^{\rm eff} (T) = R(T+1) - R(T)\,,
\eea
where
\bea
R(T) = \frac{\sum_t C_3(T,t)}{C_2(T)} \,,
\eea
The leading excited-state effects can be parameterized by
\bea\label{eq:fitform}
M^{\rm eff}(T) = M ( 1 + A e^{-\Delta T} + B T e^{-\Delta T}) \,,
\eea
where $\Delta$ is the energy gap between the ground state and the lowest excited state. This method will also have significantly reduced excited-state contamination at large Euclidean separation compared to the typical ratio method, whose excited state contamination decays as $ e^{-\Delta T/2}$. These decreased excited-state effects allow for particularly short time extents for matrix element extraction, as was demonstrated in~\cite{Chang:2018uxx}. 

For the summation method to be successful, many source-to-sink separations are required. The common sequential source technique would require a large number of propagator inversions for this to be practical. Instead, this calculation shall use a sequential operator to construct three point correlation functions. The sequential operator, $H$, is defined as the solution to the system of equations, with suppressed spin and color indices
\bea\label{eq:seqop}
\sum_{x,s}\!\! D(y,t;x,s) H(O_{\rm op};x,s;x_0,t_0) \!=\! O_{\rm op}(y,t) G(y,t;x_0,t_0),\nn
\eea
where $D(y,s;x,t)$ is the Dirac matrix and $G(y,s;x_0,t_0)$ is the point-to-all propagator from a randomly chosen source point $(x_0,t_0)$. The three point correlation function is calculated by replacing a single propagator in a standard two point function calculation with the sequential operator.

As mentioned earlier, this calculation utilizes the momentum smearing procedure~\cite{Bali:2016lva} to improve the signal of the moving states. Three values of the momentum smearing parameter $\zeta$ are used, including 0. For each of those momentum smearing parameters, two motifs of smearing are used. The source interpolating field is always smeared, but the sink quarks will be either smeared, called smeared-smeared (SS), or left as points, called smeared-point (SP). Typically the SS correlation functions will have lower excited state contamination. On the other hand, the SP correlation functions will typically have less statistical noise.

\begin{figure}[!htp]
  \centering
  \subfigure[]{\includegraphics[width=0.50\textwidth]{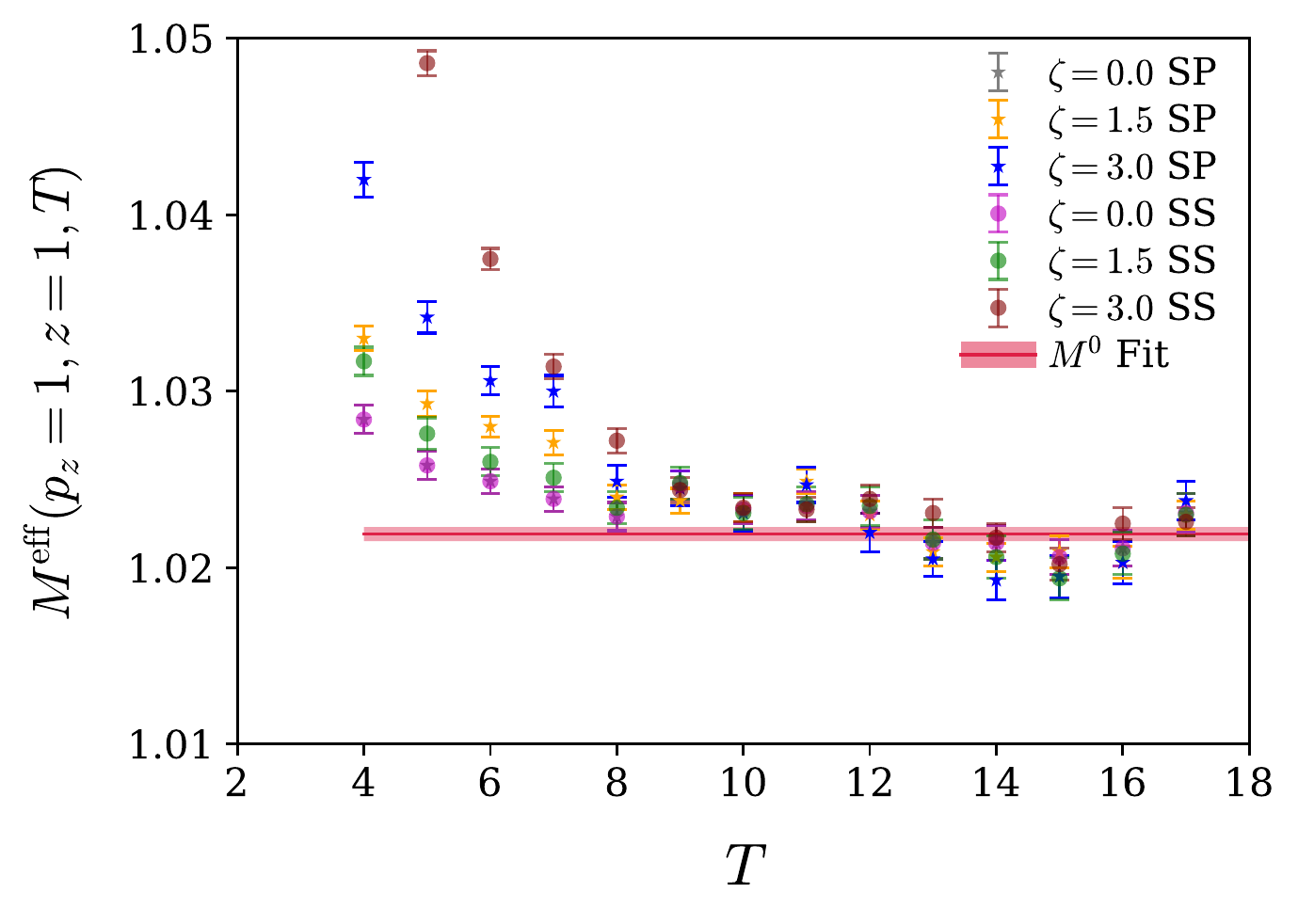}\label{2a}}
  \subfigure[]{\includegraphics[width=0.50\textwidth]{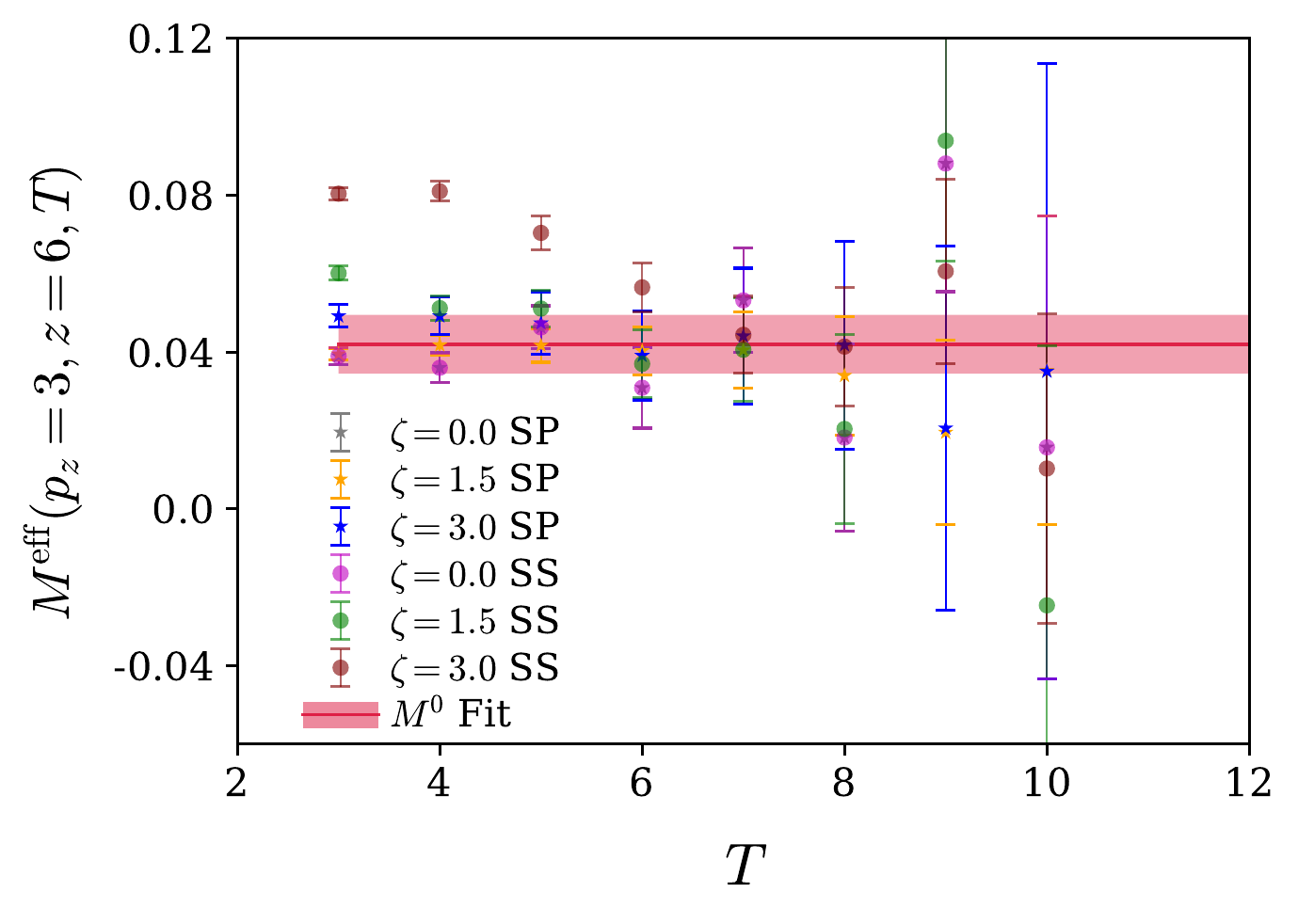}\label{2b}}
  \caption{\label{matelem} 
Example fits of the bare Ioffe time pseudo-distributions for the ensemble $a127m415L$. FIG.~\ref{2a} corresponds to the matrix element with $z=a$ and momentum $p=(2\pi)/(La)$ in the $z$-direction and denoted by $p_z=1$ in the figure. For this fit the $\chi^2$/d.o.f. is 1.17. FIG.~\ref{2b} corresponds to the matrix element with $z=6a$ and momentum $p=3(2\pi)/(La)$ in the $z$-direction and denoted by $p_z=3$ in the figure. For this fit, $\chi^2/{\rm d.o.f.} = 0.62$. The color points correspond to different correlation functions for smeared-point (SP) and smeared-smeared (SS) source and sink. Different values of momentum smearing parameters are denoted by $\zeta$. The red band corresponds to the value of the matrix element $M^0$ extracted from the fit. } 
\end{figure}

A fit of the data to Eq.~\eqref{eq:fitform} is used for each correlation function, holding the ground-state matrix element and effective energy gap fixed between each correlation function of the same $p$ and $z$. Specifically, a fit is performed on $N$ different effective bare matrix elements with different smearing setups, $M_j^{\rm eff}$, to the form

\bea
M_j^{\rm eff}(p,z, T) &=& M^0(p,z)( 1  + e^{-\Delta_p T}   [ A^{(j)}_p(z^2)  \nn \\
&+& B^{(j)}_p(z^2) T ] )
\label{eq:full_fit_form}
\eea
with $(2N+2)$ fit parameters where $j$ labels the $N$ different smearings. The fit parameters will be chosen with a weighted $\chi^2$ minimization which employs the full covariance matrix. The different momentum smearing parameters only improve the signal-to-noise ratio for a certain range of momenta states and decreases the signal-to-noise ratio for other momenta. The $T$ range  as well as which of these correlation functions are used in the fit are varied to minimize the $\chi^2$ per degree of freedom ($\chi^2$/d.o.f.). All statistical errors for the correlation functions and matrix elements are estimated using the jackknife resampling technique. Examples of these fits for the pion matrix elements are plotted in FIGs.~\ref{2a} and \ref{2b}.

\begin{figure}[!htp]
  \centering
  \subfigure[]{\includegraphics[width=0.50\textwidth]{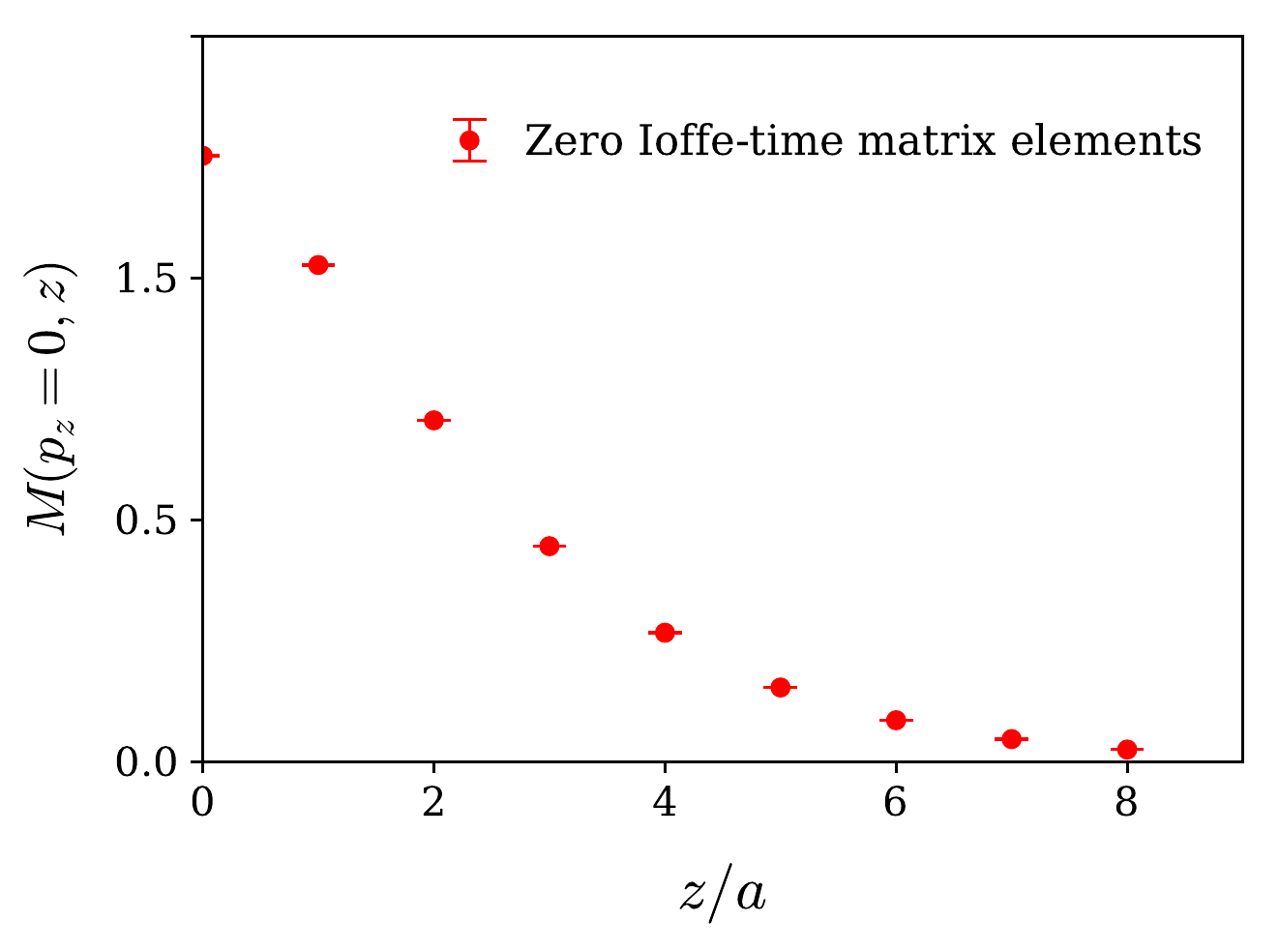}\label{2a}}
  \caption{\label{p0} 
The bare matrix element calculated for the case of $p=0$. The constant signal-to-noise ratio in $T$ allows for these pion matrix elements to be extremely precise compared to other hadrons. The exponential decay in $z$ is a feature caused by the renormalization of the Wilson line. } 
\end{figure}

The bare matrix elements for the rest frame are shown in FIG~\ref{p0}. A unique feature of pion correlation functions at zero momentum is a constant signal-to-noise ratio in $T$ allowing these matrix elements to be significantly more precise than they would be for other hadrons. The value of these matrix elements decays exponentially in $z/a$. This feature is generated by the renormalization of the Wilson line. In perturbation theory, this behavior appears as a power divergence for small $z/a$. This exponential behavior will appear in matrix elements for all momenta, but without the constant signal-to-noise ratio. The matrix elements for large distance become very small and with the exponentially growing statistical error at $p\neq 0$, they can be increasingly difficult to resolve.

In our calculation, the largest momentum along the $z$-direction we use for both of the ensembles is \mbox{$p_{\rm max} = 3(2\pi/La)$}. We use $z_{\rm max} =6a$ (0.76 fm) and \mbox{$8a$ (1 fm)} for the $a127m415$ and $a127m415L$ ensembles, respectively. Even with these relatively large separations, there does not appear a noticeable sign of higher twist effects within the limitation of present statistics. Consequently, when calculating the moments of the PDF or of the $\overline{\mbox{MS}}$ ITD, both of which can be calculated for each $z$ independently, there does not appear to be any significant dependence on the value of $z$ used as will be demonstrated in the Section~\ref{sec:mom}. If one attempts to determine the PDF from the ITD using a limited set of data, i.e. $z \leq 4a$, or using the full set of data, then the results will be consistent with each other  but with larger variance for the smaller set of data. This feature is particularly apparent in the low-$x$ region due to the shortened Ioffe time extent of the ITD.

It is important to note that the size of potential higher twist effects must be confirmed before trusting results at any separation. As is frequently done, labeling higher twist effects as $\mathcal{O}(z^2\Lambda_{\rm QCD}^2 )$ for pseudo-ITDs or $\mathcal{O}(\Lambda_{\rm QCD}^2/ p_z^2)$ for quasi-PDFs only estimates the size of these effects and the data must be checked for the presence or lack of these effects before results can be trusted. For the largest separation in this analysis, we have $z^2\Lambda_{\rm QCD}^2 \sim 1$, but the reduced pseudo-ITD appears to have successfully removed the higher twist effects in this range of Ioffe time.

\begin{figure}[!htp]
\begin{center}
\setlength\belowcaptionskip{-3pt}
\includegraphics[width=3.4in]{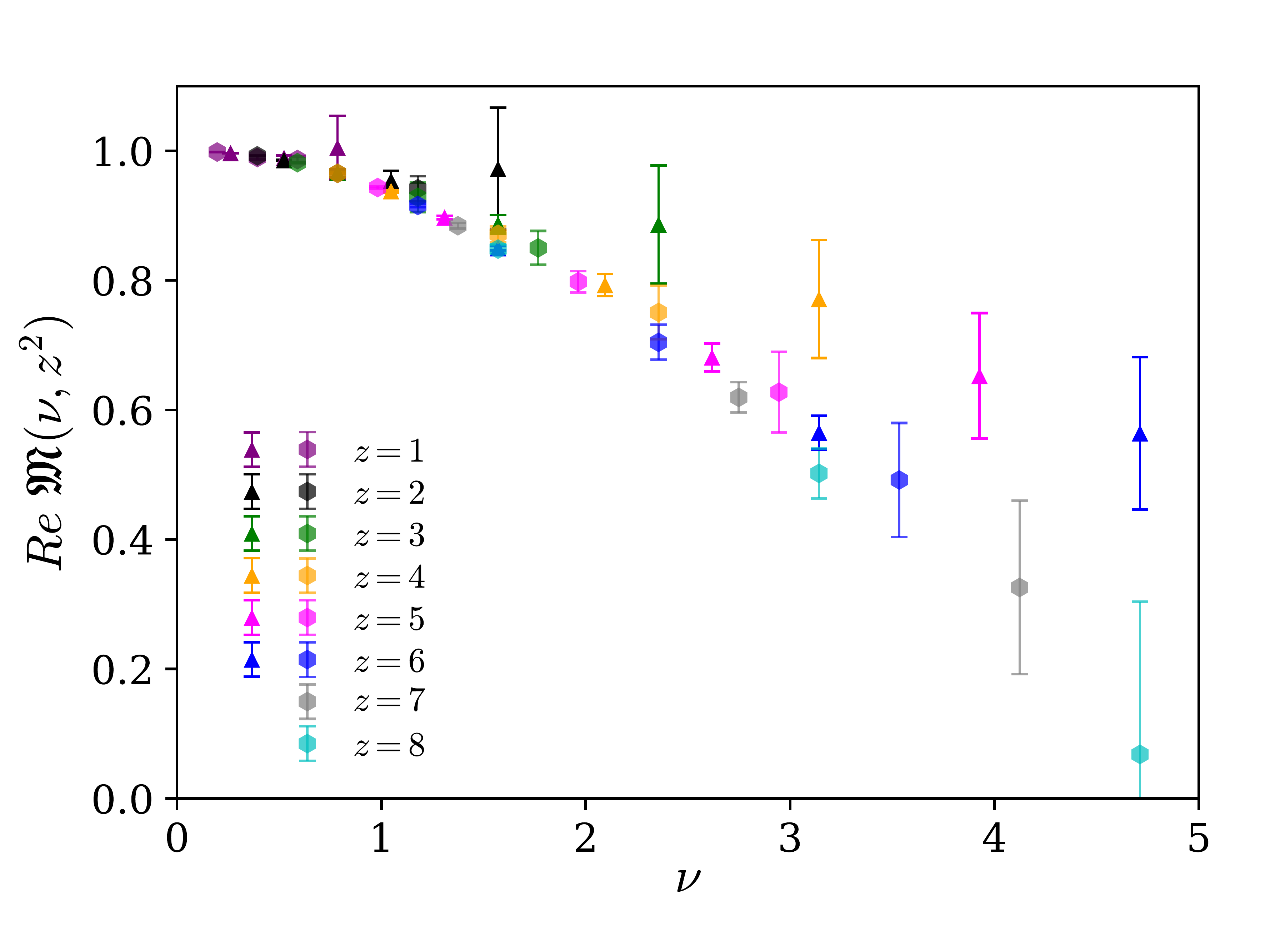}
\caption{\label{fig:matelem}
Real component of the reduced pseudo-ITD obtained from ensembles $a127m415$ and $a127m415L$ for $z_{\rm max} =6a$ and $8a$, respectively. The largest momentum used for both the ensembles is $p_{\rm max} = 3(2\pi/La)$. The triangle-up ($\triangle$) symbols indicate the reduced pseudo-ITD matrix elements $\mathfrak{M}(\nu,z^2)$ extracted from the $a127m415$ ensemble and the hexagon ($\hexagon$) symbols denote those for the $a127m415L$ ensemble. These data points represent the reduced pseudo-ITD before any perturbative matching is performed.}
\end{center}
\end{figure}

 The reduced-ITDs for both  ensembles before any perturbative matching are shown in FIG.~\ref{fig:matelem}. 
 One can see that the most data group around some $\nu$-dependent curve, with a rather small scatter.
 Excluded from this pattern are the  
  $p_z=3$ data points on the $a127m415$ ensemble, which are visibly outside  from the other data points that  follow a 
  somewhat regular distribution in $\nu$
  in FIG.~\ref{fig:matelem}. It is worth noting that while $p_z=3$ corresponds to a momentum of $\sim 0.91$ GeV in physical units on the $a127m415L$ ensemble, it corresponds to a momentum of \mbox{$\sim1.22$ GeV} on the $a127m415$ ensemble with smaller volume. In fact,  these data points are still within less than $\sim 2 \sigma$ away of the other data points, 
    and their inclusion does not affect the subsequent result of the fit to extract the pion valence PDF.

\section{Moments of the Pion PDF}
\label{sec:mom} 
As was described in~\cite{Karpie:2018zaz}, the reduced pseudo-ITD can be used to calculate the moments of the PDF. By comparing the Taylor expansions with respect to $\nu$ of Eq.~\eqref{eq:factorizable}, one can derive a multiplicative relationship between the moments of the pseudo-PDF, $b_n(z^2)$, and the moments of the $\overline{\mbox{MS}}$ PDF, $a_n(\mu^2)$
\begin{equation}\label{eq:mom_matching}
    b_n(z^2) = C_n(\mu^2z^2) a_n(\mu^2) + \mathcal{O}(z^2 \Lambda^2_{\rm QCD})
\end{equation}
where $C_n$ are the Mellin moments of the matching kernel $C(u,\mu^2z^2)$ with respect to $u$. To  next-to-leading order (NLO) accuracy, the moments are given by
\begin{equation}\label{eq:pmom_match}
C_n(z^2\mu^2) =  1  -  \frac{\alpha_s}{2\pi} C_F \left[\gamma_n \ln\left(z^2\mu^2\frac{e^{2\gamma_E +1}}{4}\right) + l_n\right]\,,
 \end{equation}
where
\begin{equation} 
\gamma_n = \int_0^1 du\, B(u) u^n= \frac{1}{(n+1) (n+2) } - \frac{1}{2}  
- 2 \sum_{k=2}^{n+1}\frac{1}{k}\,, 
\end{equation}
 are the well known moments of the Altarelli-Parisi kernel, and
 \begin{align}
l_n = \int_0^1 du\, L(u) u^n=&2\left[ \left(\sum_{k=1}^n \frac{1}{k}\right)^2 + \sum_{k=1}^n \frac{1}{k^2}
\right. \nn \\ & \left.  
+\frac12 - \frac{1}{(n+1)(n+2)} \right]
\, .
\end{align}

By completely avoiding the inverse problem~\cite{Karpie:2018zaz}, this procedure allows for an understanding of the PDF's structure before any potential systematic errors arising from the matching convolution and Fourier transforms are incurred. In principle, this method, coined as ``OPE without OPE''~\cite{Martinelli:1998hz}, can be used to determine any moment of the PDF in sharp contrast to the traditional method which is based on local matrix elements. The latter is limited by the appearance of power divergent mixing due to the reduced rotational symmetry of the lattice as well as issues related to the signal-to-noise ratio. In practice, however, only the lower moments will have a resolvable signal. Larger Ioffe time extents and a finer resolution in Ioffe time, both of which require finer lattice spacings, are necessary before higher moments can be obtained.

\begin{figure}[h]
\begin{center}
\setlength\belowcaptionskip{-3pt}
\includegraphics[width=3.4in]{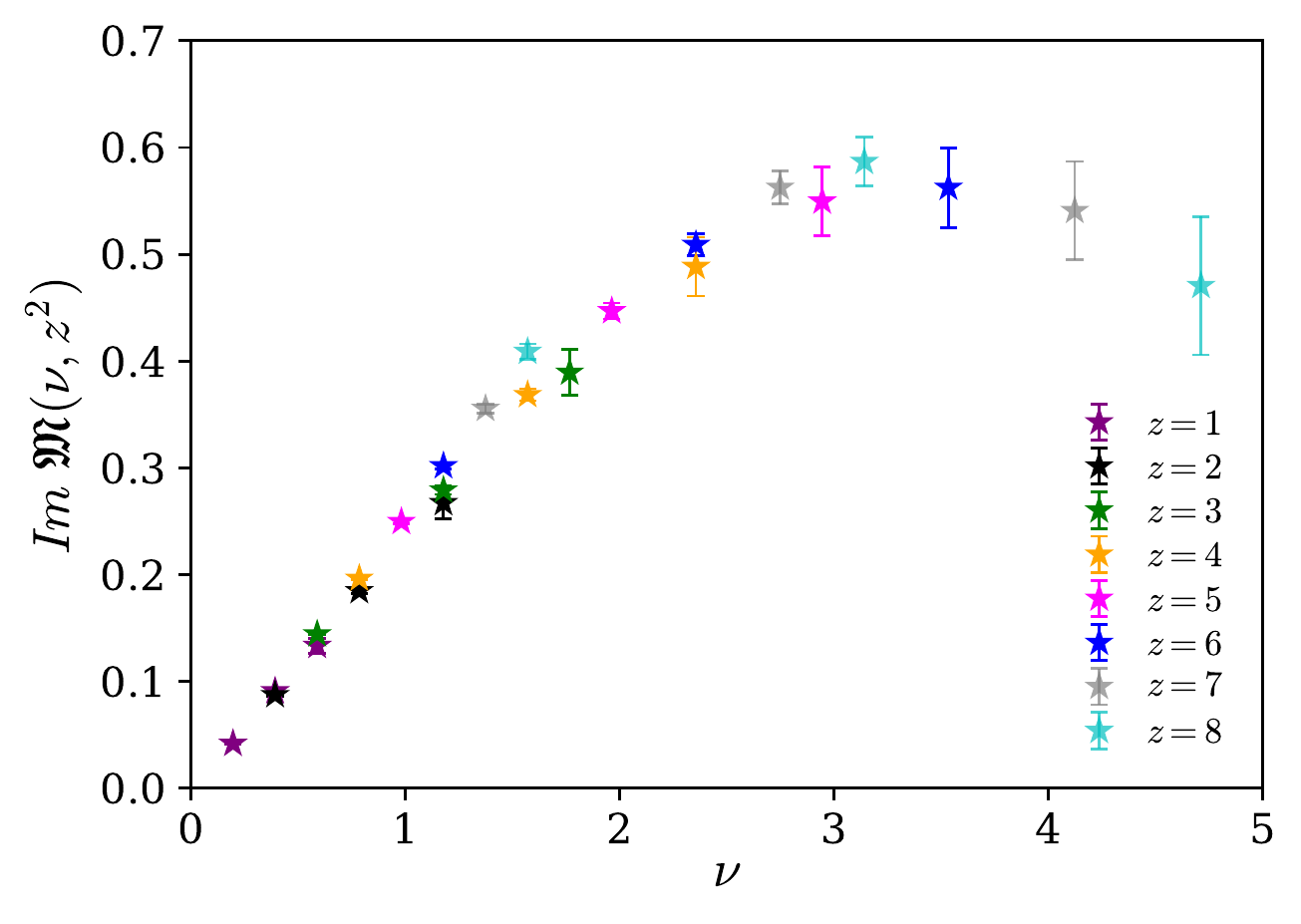}
\caption{\label{fig:ImITD32}
Imaginary component of the reduced pseudo-ITD obtained from the ensemble $a127m415L$ for $z_{\rm max}=8a$. The largest momentum is $p_{\rm max} = 3(2\pi/La)$.  }
\end{center}
\end{figure}

The lowest four moments will be extracted from the $a127m415L$ ensemble data by inverting the Vandermonde matrix as was performed in~\cite{Karpie:2018zaz,Joo:2019jct}. The imaginary  and real components are used individually to calculate the odd and even moments respectively. The imaginary component of the reduced pseudo-ITD obtained from the $a127m415L$ ensemble is shown in FIG.~\ref{fig:ImITD32}. Due to the larger uncertainty of the $p_z=3$ data points from the $a127m415$ ensemble and the deviation from the Ioffe time distribution of the other data points as shown in FIG.~\ref{fig:matelem}, this ensemble does not allow us to extract moments in a reliable way. Therefore, we extract moments using the real and imaginary components of the reduced-ITD only from the $a127m415L$ ensemble in what follows.

\begin{figure}[h]
\begin{center}
\includegraphics[width=3.4in]{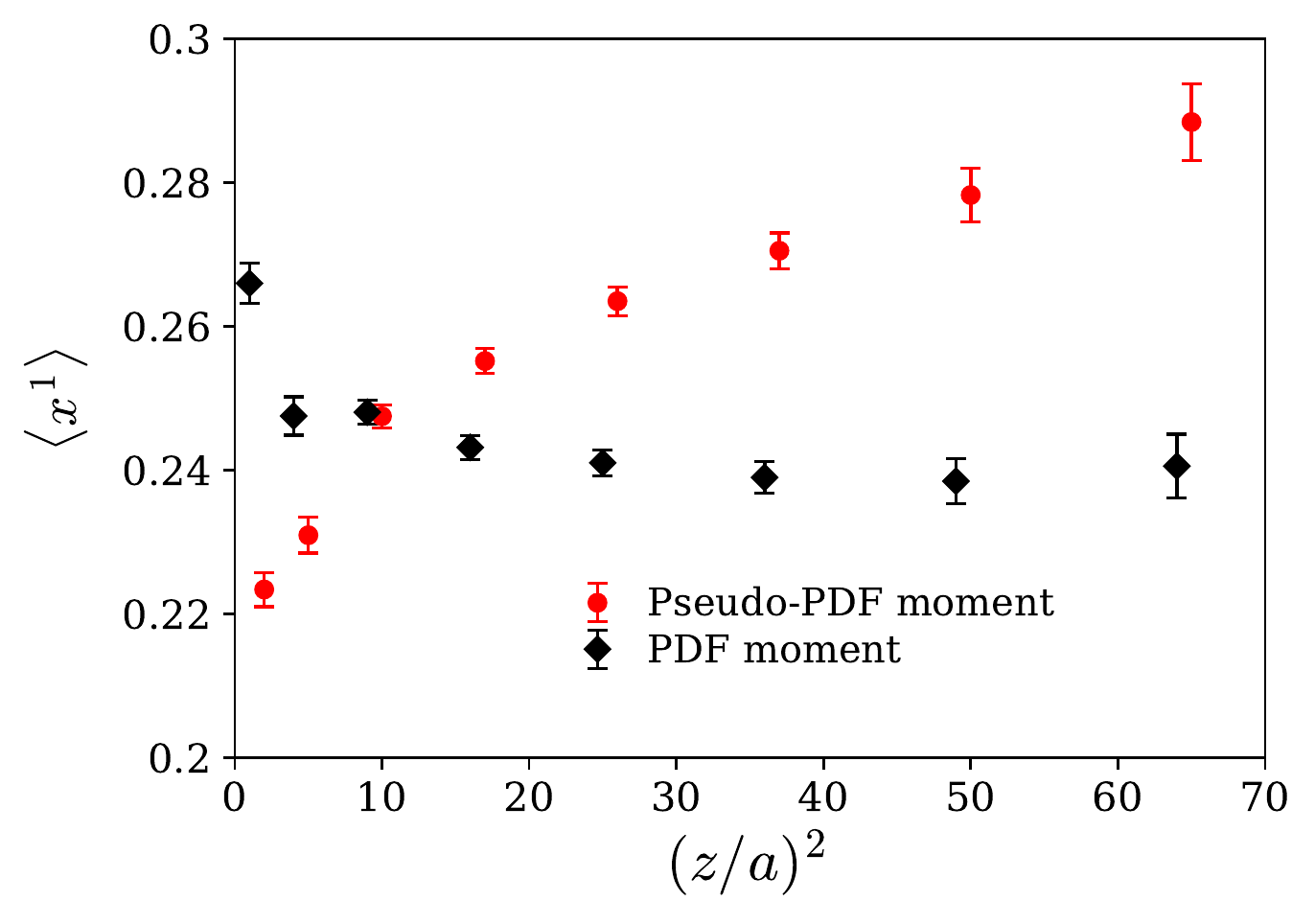}
\caption{\label{fig:mom1}
The first moments of the pion pseudo-PDF and of the PDF calculated  from the $a127m415L$ ensemble are shown. The moments are shown as functions of the original separation $(z/a)^2$ of the reduced pseudo-ITD data used to calculate them with small offsets for better visibility. After the matching procedure, the dependence on the separation is significantly reduced. The largest deviation occurs for the lowest separation, which is most susceptible to discretization errors. }
\end{center}
\end{figure}

\begin{figure}[!htp]
\begin{center}
\includegraphics[width=3.4in]{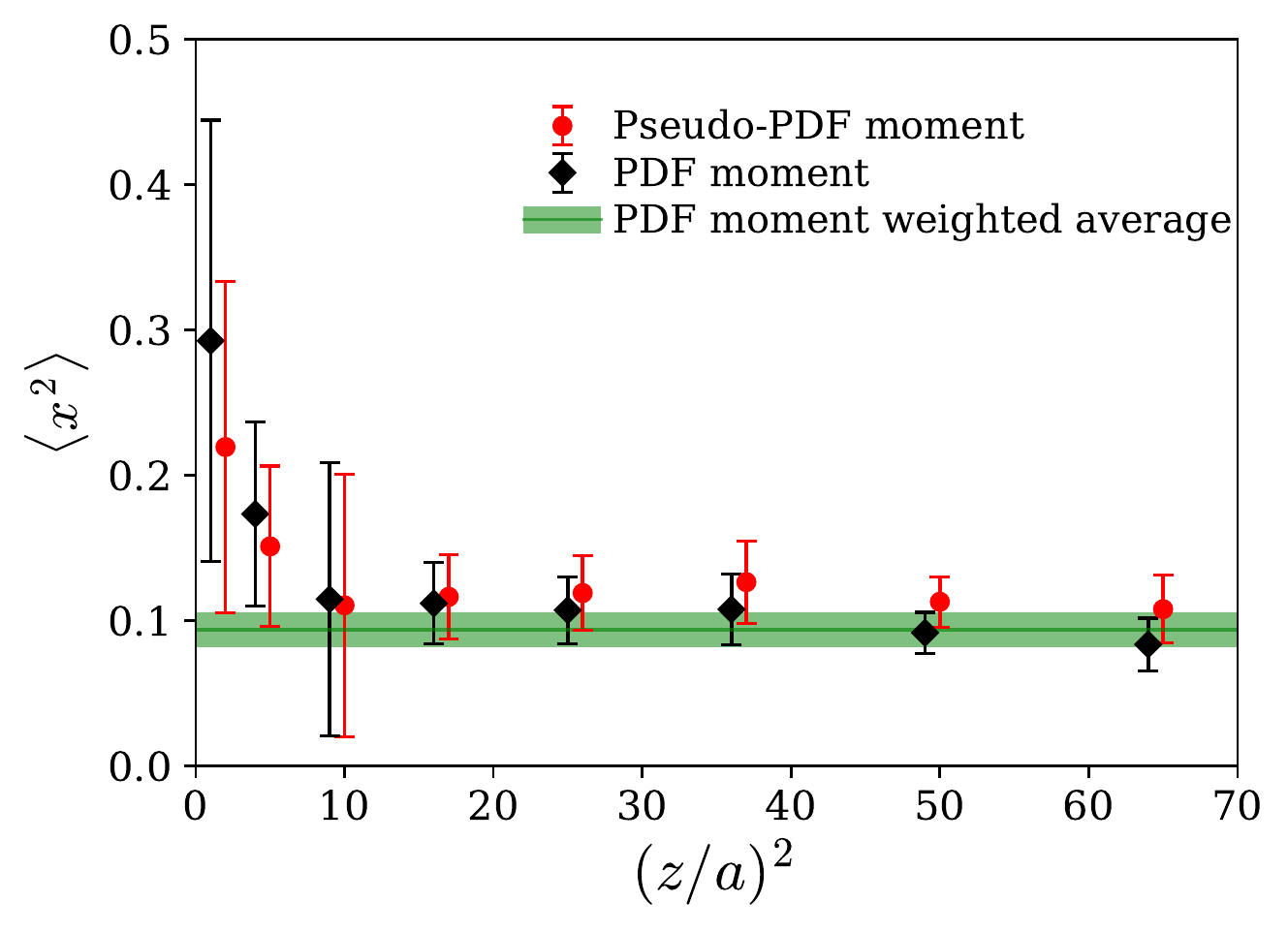}
\caption{\label{fig:mom2}
The second moments of the pion pseudo-PDF and of the PDF calculated from the  $a127m415L$ ensemble are shown. The moments are shown as functions of the original separation $(z/a)^2$ of the reduced pseudo-ITD data used to calculate them with small offsets for better visibility. After the matching procedure, the dependence on the separation vanishes within statistical precision, showing the lack of higher twist effects in this moment at this level of precision. The green band represents the PDF moment from a weighted average. 
}
\end{center}
\end{figure}

\begin{figure}[!htp]
\begin{center}
\includegraphics[width=3.4in]{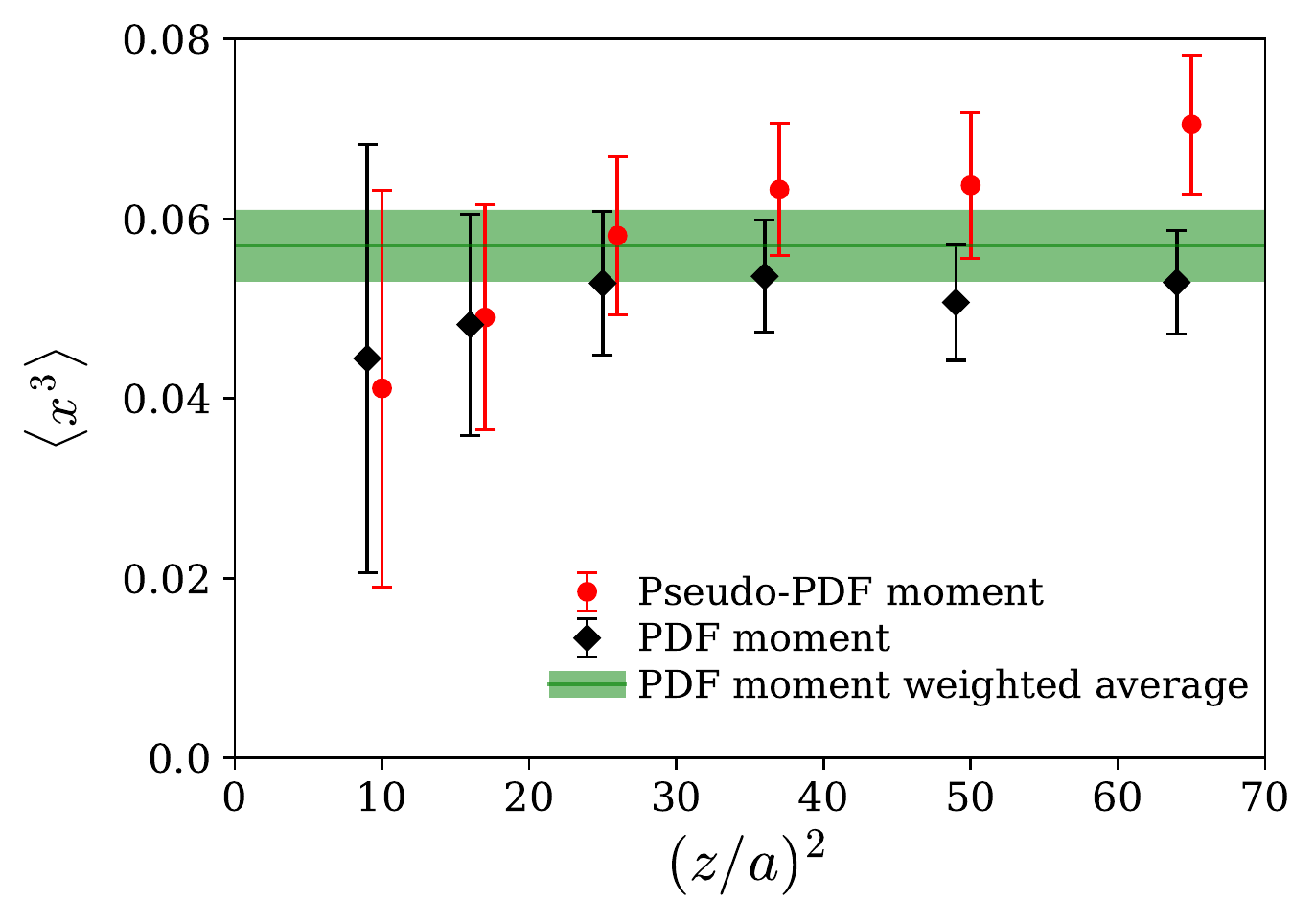}
\caption{\label{fig:mom3}
The third moments of the pion pseudo-PDF and of the PDF calculated from the $a127m415L$ ensemble are shown. The moments are shown as functions of the original separation $(z/a)^2$ of the reduced pseudo-ITD data used to calculate them with small offsets for better visibility. The data from lowest two $z$ values in this calculation only show linear effects due to the short range of Ioffe time they span. As a result, they do not have any signal for the third moment and are not shown here. After the matching procedure, the dependence on the separation vanishes within statistical precision, showing the lack of large higher twist effects at this level of precision. The green band represents the PDF moment from a weighted average. }
\end{center}
\end{figure}

\begin{figure}[!htp]
\begin{center}
\includegraphics[width=3.4in]{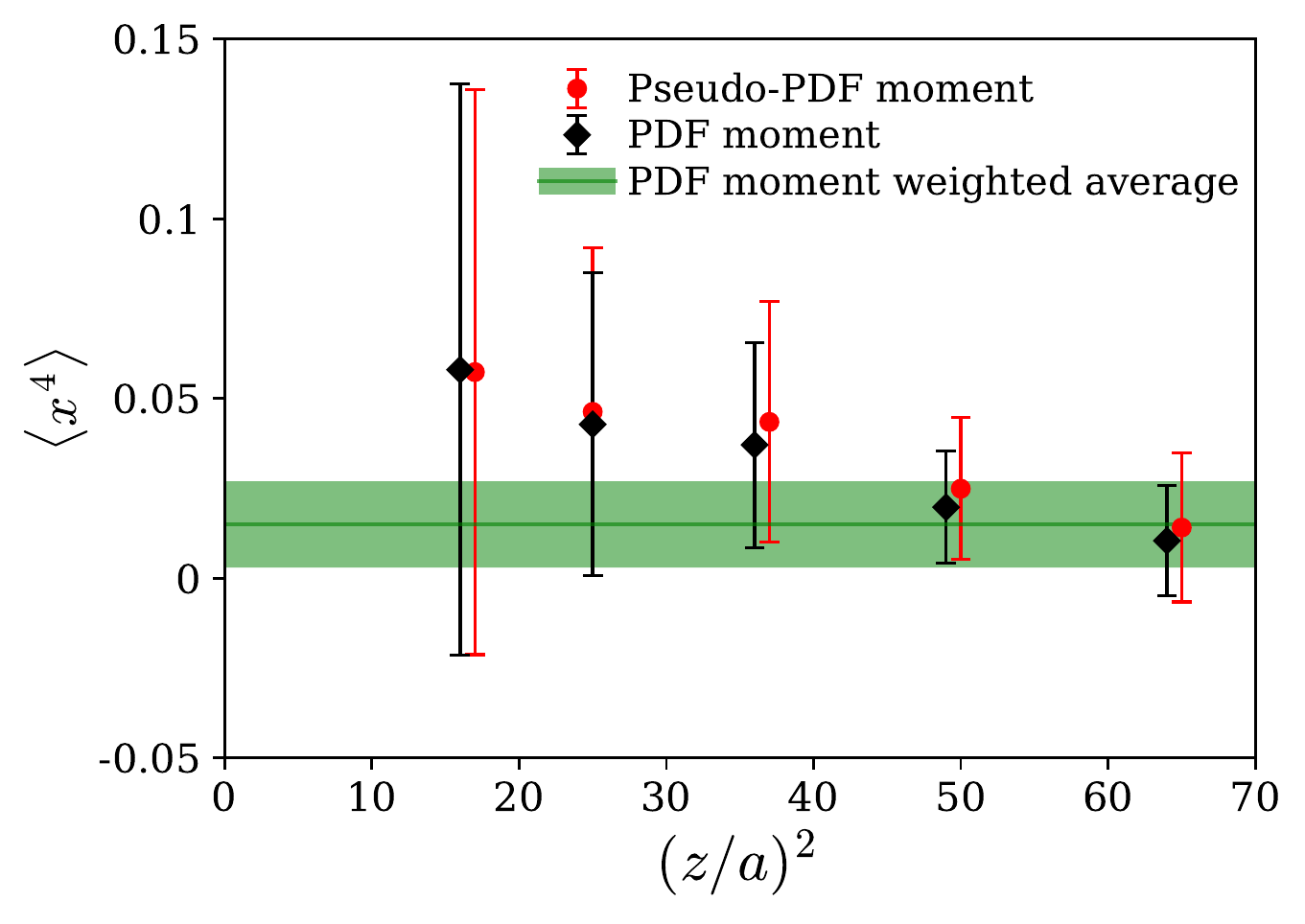}
\caption{\label{fig:mom4}
The fourth moments of the pion pseudo-PDF and of the PDF calculated from the $a127m415L$ ensemble are shown. The moments are shown as functions of the original separation $(z/a)^2$ of the reduced pseudo-ITD data used to calculate them with small offsets for better visibility. The data from the lowest three $z$ values in this calculation only show linear effects due to the short range of Ioffe time they span. As a result, they do not have any signal for the fourth moment and are not shown here. After the matching procedure, the dependence on the separation vanishes within statistical precision, showing the lack of large higher twist effects at this level of precision. The green band represents the PDF moment from a weighted average. }
\end{center}
\end{figure}

The results for these moments, determined from data with different $z$ independently, are shown in FIGs.~\ref{fig:mom1},~\ref{fig:mom2},~\ref{fig:mom3}, and~\ref{fig:mom4}, each one called ``pseudo-PDF moment" before matching and ``PDF moment" after matching. For the matching relationships, we choose $\mu = 2$ GeV, and  the value of $\alpha_s(2\,{\rm GeV})=0.303$. This value of the coupling is taken from the evolution used by the LHAPDF~\cite{Buckley:2014ana} for the dataset cj15nlo from the CTEQ-Jefferson Lab collaboration~\cite{Accardi:2016qay}. The imaginary component of the pseudo-ITD calculated with the lowest two $z$-values has a completely linear behavior within their short Ioffe time range. This  leads to an inaccurate determination of the third moment, which also slightly affected the calculation of the first moment. The first moment for these two separations is instead calculated with a linear fit. Similarly, the real component of the reduced pseudo-ITD calculated with the lowest three $z$-values only exhibits a quadratic behavior and has an ill-constrained value for the fourth moment. In this case, those results did not appear to affect the quality of the second moment determination, so the results are kept for the second moment. For the third and fourth moments, these poorly constrained results are dropped from the following analysis.


 After the matching procedure in Eq.~\eqref{eq:mom_matching} has been applied, there is no significant large-$z$ dependence on the separation $z$ from which the data originated for the second, third, and fourth moments. On the other hand, a $z^2$-dependence does appear for the first moment calculation, particularly between the $z=1a$ and $1a<z\leq 8a$ calculations. This moment at $z=1a$ point is the most sensitive to the lattice spacing errors. In addition, this point was determined with a simple linear fit and therefore it can be affected by different systematics.  
 
  \begin{table*}[!htp]
  \centering
  \begin{tabular}{| c | c | c | c | c | c | c |}
  \hline
    fit  & $a$ & $c_2$  & $c_{2l}$ &  $c_4$ & $c_{4l}$ & $\chi^2/{\rm d.o.f}$\\
    \hhline{|=|=|=|=|=|=|=|}
    $a_2^{\rm latt}$ & $0.2478(15)$ & -0.0054(16) & - & - & - &  5.42 \\
    $a_{2l}^{\rm latt}$ & $0.2515(17)$ & -0.051(17) & 0.008(3) & - & - &  1.7 \\
    $a_4^{\rm latt}$ & $0.2541(26)$ & -0.015(3) & - & 0.004(1) & - &  1.2 \\
   $a_{4l}^{\rm latt}$ & $0.239(57)$ & 0.2(1.2) & -0.07(35) & 0.3(1.4) & -0.03(21) &  0.67 \\
    \hhline{|=|=|=|=|=|=|=|}
  \end{tabular}
  \caption{The parameters from fitting the residual $z^2$ dependence of the first PDF moment for the $a127m415L$ ensemble. The higher twist terms are smaller than what would have been been expected by a simple $O(z^2\Lambda^2_{\rm QCD})$ estimate. }
  \label{tab:ht_fits}
\end{table*}
 
\begin{figure}[!htb]
\begin{center}
\setlength\belowcaptionskip{-3pt}
\includegraphics[width=3.4in]{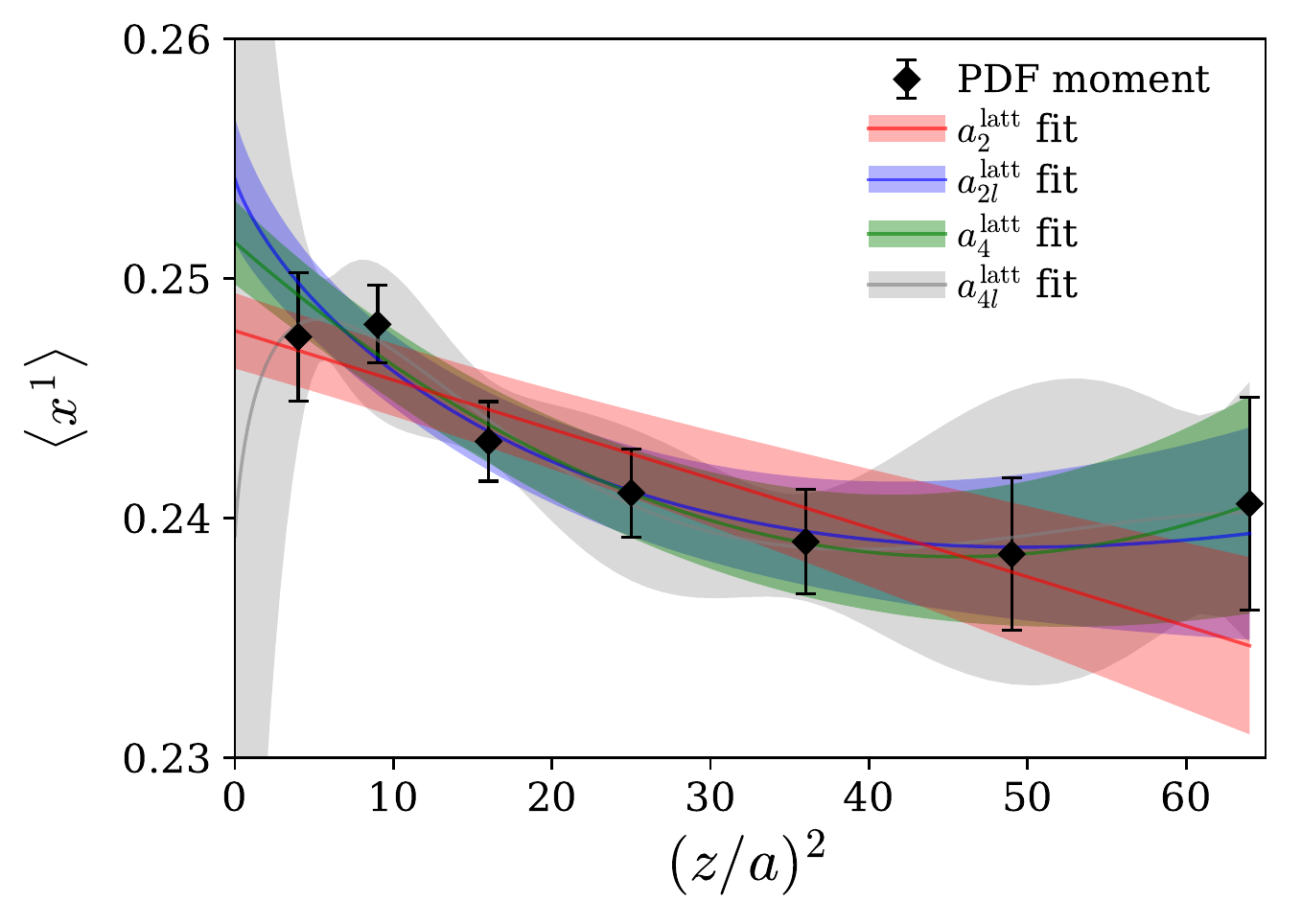}
\caption{\label{fig:ht_fits}
The first moment of the PDF calculated from the $a127m415L$ ensemble is  shown. The different bands correspond to different models of the higher twist effects. These effects are significantly smaller than what an $O(z^2 \Lambda^2_{\rm QCD})$ estimation would have provided. The value $<x^1>=0.2541(26)$ denoted by the green band in $a_4^{\rm latt}$ fit is quoted in Table~\ref{tab:mom}.}
\end{center}
\end{figure}

For the first moment  the residual z-dependence at larger $(z/a)^2$, which may be due to  higher twist effects can be modeled. These effects are either polynomial \mbox{(i.e. $z^{2n}$)} or of the form $z^{2n} \ln(z^2\mu^2\frac{e^{2\gamma_E +1}}{4})$. 
The logarithmic terms arise from logarithms in the matching coefficients $C_n(z^2\mu^2) $ when applied to pseudo-PDF moments that contain higher twist effects. As we can see these terms are suppressed by $\alpha_s$ and they are expected to be smaller than the simple polynomial terms.  Due to its large potential discretization errors, the $z=a$ datapoint is neglected in this fit. The moments are fit to four different functional forms
 \bea\label{moment_fits}
 a_2^{\rm latt}(z^2) = a + c_2 z^2\Lambda_{\rm QCD}^2 \,, \nn\\
a_{2l}^{\rm latt}(z^2) = a + c_2 z^2\Lambda_{\rm QCD}^2 + c_{2l} z^2\Lambda_{\rm QCD}^2 \ln(z^2\mu^2\frac{e^{2\gamma_E +1}}{4}) \,, \nn\\
a_4^{\rm latt}(z^2) = a + c_2 z^2\Lambda_{\rm QCD}^2 + c_4 z^4\Lambda_{\rm QCD}^4 \,,\nn \\
a_{4l}^{\rm latt}(z^2) = a + c_2 z^2\Lambda_{\rm QCD}^2  + c_{2l} z^2\Lambda_{\rm QCD}^2 \ln(z^2\mu^2\frac{e^{2\gamma_E +1}}{4}) \nn\\+ c_4 z^4\Lambda_{\rm QCD}^4  + c_{4l} z^4\Lambda_{\rm QCD}^4 \ln(z^2\mu^2\frac{e^{2\gamma_E +1}}{4}) \nn \,,
 \eea
 and the results of these fits are shown in Table~\ref{tab:ht_fits} and are plotted in FIG.~\ref{fig:ht_fits}. The value of $\Lambda_{\rm QCD} = 300$ MeV was used, but this choice is made solely for the magnitude of the coefficients of the $O(z^2 \Lambda^2_{\rm QCD})$ terms to be estimated. From the value of the coefficients in Table~\ref{tab:ht_fits}, the higher twist effects in this moment are an order of magnitude smaller than  what a na\"ive $O(z^2 \Lambda^2_{\rm QCD})$ estimate would have suggested. The first moment of the PDF dominates the low $\nu$ behavior of the pseudo-ITD and ITD, which is precisely the region where the reduced pseudo-ITD reduces higher twist effects. The systematic error introduced by these higher twist effects would be comparable to, or smaller than, the other systematic errors of finite lattice spacing and unphysical pion mass.
If there had appeared a sufficiently strong polynomial $z^2$-behavior in enough moments, one could use the inverse Mellin transform of that behavior to estimate the higher twist effects in the reduced pseudo-ITD. With only four moments and only one of them showing any discernible higher twist effects, this inverse transform will not be reliable. This lack of large higher twist effects in the moments, even with separations as large as $z=8a$, justifies the use of this data for extracting the PDF. 

\begin{table}[!htb]
  \centering
  \begin{tabular}{| c | c | c |}
  \hline
 n & $\langle x^n \rangle$ & $z_{\rm min}$ \\
    \hhline{|=|=|=|}
    1 & 0.2541(26) & 2a \\
    2 & 0.094(12) & 1a \\
    3 & 0.057(4)  & 3a \\
    4 & 0.015(12)  & 4a \\
    \hhline{|=|=|=|}
  \end{tabular}
  \caption{The moments of the PDF determined on the $a127m415L$
      ensemble through the ``OPE without OPE'' method, and the lowest
    separations used in the calculation.   }
  \label{tab:mom}
\end{table}

We summarize our calculation of the moments of the pion PDF in Table~\ref{tab:mom}. For the moments which do not show signs of higher twist effects, we take a covariance weighted average of the results for each separation. For the first moment, we state the value from the fit $a_4^{\rm latt}$ with the higher twist contamination removed. The fit $a_{4l}^{\rm latt}$ has large values of the variance of the fit parameters, a small number of degrees of freedom, and the oscillatory nature of the final result. We believe this result obtained from from $a_{4l}^{\rm latt}$ is overfit. We
refer the readers to the previous
calculations~\cite{Best:1997qp,Guagnelli:2004ga,Capitani:2005jp,Detmold:2005gg,Bali:2013gya,Abdel-Rehim:2015owa,Oehm:2018jvm}
of such moments for a comparison with the calculated moments in this work. 

We shall calculate the moments from the extracted pion valence PDF, in the next section, and compare with those obtained from in the NLO QCD analysis~\cite{Wijesooriya:2005ir} from the Fermilab E-615 pionic Drell-Yan data~\cite{Conway:1989fs} at a scale of 5.2 GeV. Note that the odd moments listed in Table~\ref{tab:mom}, which are extracted from the imaginary component of the reduced pseudo-ITD, are related to $q_{\rm v}(x)+2\bar{q}(x)$ distribution as shown in Ref.~\cite{Orginos:2017kos}. Therefore, our results for the odd moments are not directly comparable with that calculation.  

\section{Extraction of the Pion Valence Distribution} 
\label{results}

It can be seen from FIG.~\ref{fig:matelem} that the $z^2$-dependence  indeed cancels out to a great extent in the reduced pseudo-ITD of Eq.~\eqref{ratio} without spoiling the $\nu$-dependence in the Ioffe time distribution which governs the shape of the pion PDF.  As discussed earlier, for small $z^2$  the function ${\mathfrak M}(\nu,z^2)$
should contain  $\ln (z^2)$ singularities related to the  perturbative evolution of the PDFs. 
Such a logarithmic $z^2$-dependence is clearly seen in the  $z\leq 4a$  data of Ref.~\cite{Orginos:2017kos,Karpie:2017bzm}, performed in the
  quenched approximation,  though at a finer lattice spacing $a \simeq
  0.093~{\rm fm}$. For large $z\geq 6a$ values, they  found the data in practice did not depend on $z$.
Thus, one can explicitly identify  the region $z\leq 4a$, where one may rely on the perturbative evolution.
As   demonstrated in~\cite{Radyushkin:2018cvn}, the   matching procedure applied in this region to the  points  
for ${\mathfrak M}(\nu,z^2)$ converts their  $\ln (z^2)$-dependence   
 into the $\ln (\mu^2)$-dependence 
of the $\overline{\rm MS}$ light-cone ITDs   $Q(\nu,\mu)$
 on the $\overline{\rm MS}$ scheme subtraction parameter $\mu^2$.  
Just like in Ref.~\cite{Joo:2019jct}, for  the matching of the reduced pseudo-ITD to the 
$\overline{\rm MS}$ ITD at a particular scale $\mu$, we perform an inversion of Eq.~\eqref{evolution} simply by switching $Q(\nu,\mu)$ and $\mathfrak{M}(\nu,z^2)$ and changing the sign of $\alpha_s$. This gives
\bea \la{matching}
Q(\nu,\mu) &=& \mathfrak{M}(\nu,z^2) - \frac{\alpha_s C_F}{2\pi} \int_0^1 du \bigg[\ln \bigg( z^2\mu^2\frac{e^{2\gamma_E+1}}{4}\bigg)\times \nn \\
&&B(u)+L(u)\bigg] \mathfrak{M}(u\nu,z^2).
\eea
In other words,  applying   the matching procedure for an appropriate  value of $\alpha_s$, the  
$z^2$-dependence of the original small-$z^2$  data for $ \mathfrak{M}(\nu,z^2)$  should be compensated by the $\ln z^2$ term,
and one should get practically $z^2$-independent data points for $Q(\nu,\mu)$.

In the present calculation, due to significantly larger uncertainties compared to those in~\cite{Orginos:2017kos}, 
no systematic  logarithmic dependence of the data  on  \mbox{$z^2$}  
is visible. Therefore, it   is not possible  to  determine  what is  the length scale $z_0$ 
below which one may rely on   perturbative evolution of the reduced pseudo-ITD,
and what is the value of $\alpha_s$ associated with this evolution.
As   guidance for a 
 particular choice of $\alpha_s$  one may use the idea that using  the correct choice of $\alpha_s$ 
 in the matching formula~\eqref{matching} 
 should produce the least scatter for the  small-$z^2$ points from a 
  universal curve  in $\nu$, at least up to discretization effects.


%
\begin{figure}[!htp]
\begin{center}
\includegraphics[width=3.4in]{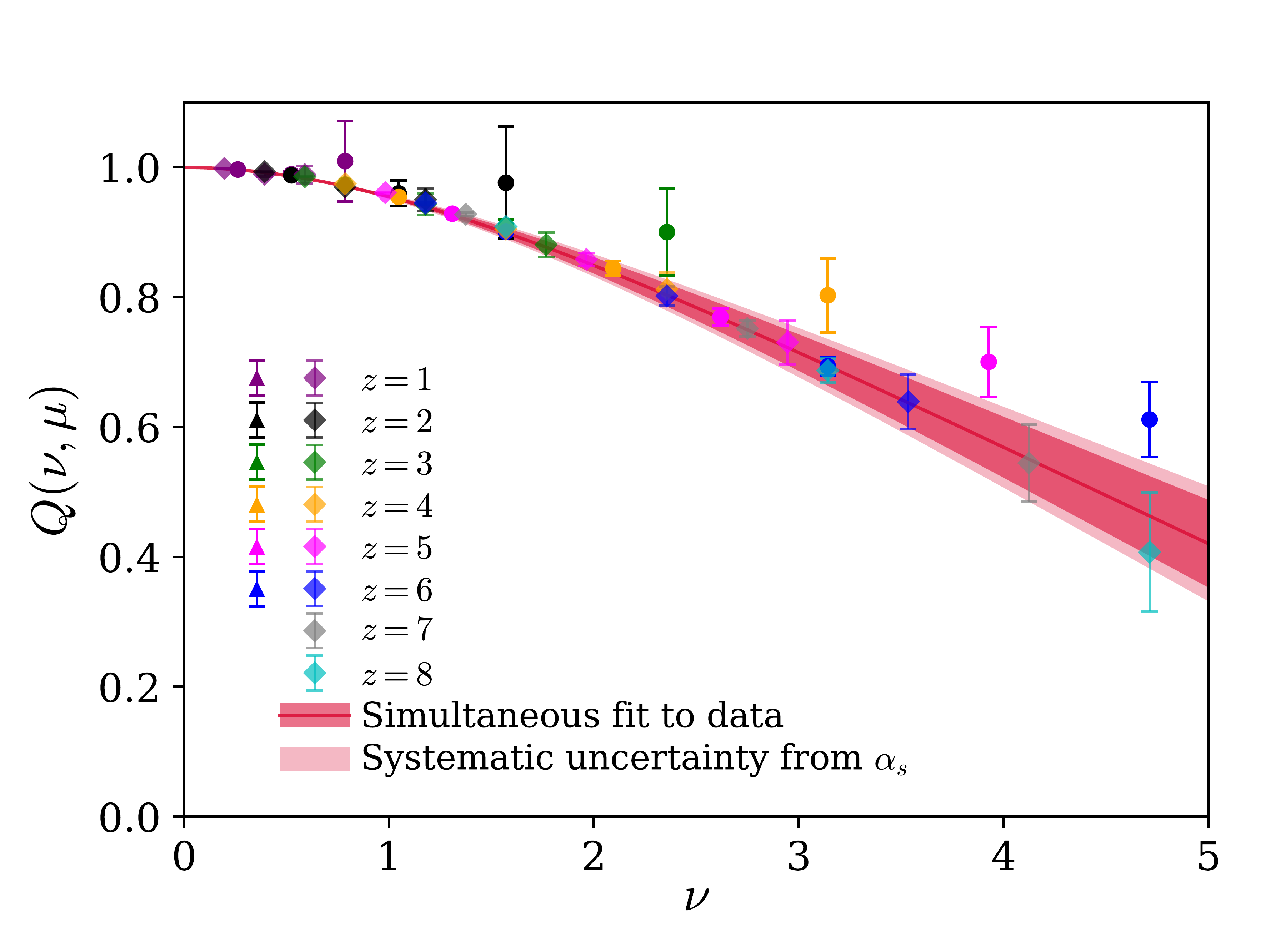}
\caption{\label{fig:mathced} The $\overline{\rm MS}$ Ioffe time distribution obtained from ensembles $a127m415$ and $a127m415L$ after 1-loop perturbative matching using Eq.~\eqref{matching} at $\mu=2$ GeV. The circle ($\circ$) symbols indicate the reduced pseudo-ITD matrix elements $M^0$ extracted from the $a127m415$ ensemble and the diamond ($\diamond$) symbols denote those for the $a127m415L$ ensemble. The  inner red uncertainty band is obtained from a simultaneous fit to the matched ITDs at $\mu=2.0$ GeV on these two ensembles in the limit of infinite volume. The outer red band is an estimate of the systematic uncertainty arising from the choice of scales obtained through a simultaneous fit to the matched ITDs in which the value of $\alpha_s$ is varied by 10\% about the central value of $\alpha_s = 0.303$ at $\mu=2$ GeV.  }
\end{center}
\end{figure}

The convolution is performed on a polynomial fit of  the data for  $ \mathfrak{M}(u\nu,z^2)$,  and Eq.~\eqref{matching} is applied for each $z$ independently. We choose $\mu = 2$ GeV, and  the value of $\alpha_s(2\,{\rm GeV})=0.303$ has been taken form the evolution used in Ref.~\cite{Buckley:2014ana} as mentioned earlier. It important to note that,  in the reduced-ITD approach, the relevant scale for converting the matrix element to the $\overline{\rm MS}$ scheme is the separation between the quark fields $z^2$, not the hadron's momentum $p_z$. To investigate the systematics of the one-loop matching on our choice of $\alpha_s$ at a particular scale, we vary $\alpha_s$ by $10\%$, and estimate its effect as a source of systematic uncertainty, as shown in FIG.~\ref{fig:mathced}. One can show that only the data points at large $\nu$ have as large as 5\% change in their central values, and are still statistically consistent between different choices of $\alpha_s$. Other points at low Ioffe time have less than 1\% change for the variation in $\alpha_s=0.303\pm0.030$. The small differences in the matched reduced-ITDs of the two ensembles originating from 10\% change in $\alpha_s$ are propagated as systematic uncertainty in the subsequent analyses. The matched ITDs for the two ensembles at the matching scale of 2 GeV are shown in FIG.~\ref{fig:mathced}. 
The matched data points in FIG.~\ref{fig:mathced} are also seen to be located  along a single curve that is  a function of $\nu$,
with rather small    fluctuations.  Again, the exceptions are 
the  
  $p_z=3$ data points from  the $a127m415$ ensemble.
 
Having data from two ensembles that have different  volumes, one may wish to use them to study
effects due to finite volume. We do not see any finite volume effect in the reduced-ITDs with the exception of $p_z=3$ data obtained from the $a127m415$ ensemble which is not trustworthy for the small volume. Moreover,  the present case of only two-volumes is not an ideal situation for an infinite-volume extrapolation. In our particular case, there are few pairs of data from the two ensembles that correspond to the same Ioffe time $\nu$.  

To extract a single PDF combining the results from the two ensembles, we will  perform a  simultaneous and correlated fit to these two data sets. The number of configurations of the two ensembles we use are different. Therefore, to perform a simultaneous fit of the matched Ioffe time distributions, equal number of bootstrap samples are generated from the two ensembles. Because of the reason mentioned earlier, we exclude the $p_z=3$ data from  the $a127m415$ ensemble in this fit.  
Because the functional form of the ITD $Q(\nu)$ is not known {\it a priori}, we implement a \mbox{``$z$-expansion"~\cite{Boyd:1994tt,Bourrely:2008za}} 
 fit to the data sets, reflecting the analyticity of $Q(\nu)$.  We also investigate whether there is residual $z^2$-dependence in the matched  $Q(\nu,\mu)$ 
 distribution. This is achieved  by adding $z^2$-dependent terms in the fit function. Finally, a term which describes a potential volume dependence is added. There are no model calculations for the form of the finite volume corrections to the matrix element in Eq.~\eqref{eq:matelem}, unlike the two current matrix element~\cite{Briceno:2018lfj}. Instead, we take a simple exponential volume dependence where the relevant distance is the difference between the lattice size $L$ and the length of the Wilson line $z$. Namely,  we use  the following form
\bea \la{zfit}
Q(\nu,z^2) &=& \sum_{k=0}^{k_{\rm max}} \lambda_k \tau^k \times \big[1+ \nu^2\big(c_1 z^2 +\nn \\
&& c_2 z^2 \ln \bigg(z^2\mu^2\frac{e^{2\gamma_E+1}}{4}\bigg) + c_3 e^{-m_\pi (L-z)}\big)\big]\nn \\
\eea
where
\bea \la{cut}
\tau = \frac{\sqrt{\nu_{\rm cut}+\nu}-\sqrt{\nu_{\rm cut}}}{\sqrt{\nu_{\rm cut}+\nu}+\sqrt{\nu_{\rm cut}}}.
\eea

Note that, unlike the form factors we do not have cuts in the complex plane for the calculated reduced pseudo-ITD and we simply choose a dimensionless number \mbox{$\nu_{\rm cut}=1.0$} in Eq.~\eqref{zfit}. In fact, other choices are possible with the final results being unaffected. 

One can readily see from Eq.~\eqref{cut} that, for fixed values of $\lambda_k$, a larger $\nu_{\rm cut}$ dictates a slower fall-off of the distribution. However, if one allows to vary $\lambda_k$ in the fit along with different choices of $\nu_{\rm cut}$, the fit parameters are also changed accordingly such that the red band shown in FIG.~\ref{fig:mathced} remains unchanged.   The value  $\lambda_0=1.0$ in the fit is fixed by the normalization in Eq.~\eqref{ratio} at $\nu=0$ and does not change by the perturbative matching. We limit the $k_{\rm max}$ in our fit to the value in which additional term in the $z$-expansion has no effects on the fit and obtain $k_{\rm max}=4$. The fit parameters are listed in Table~\ref{tab:zfit}. The smallness of the fit parameters $c_1$ and $c_2$ reflect the fact that the residual $z^2$-dependence is negligible as discussed earlier. The red band in FIG.~\ref{fig:mathced} represents the ITD in the limit of infinite volume and vanishing $z^2$-contribution. The outer red band indicates the systematic uncertainty in the $z$-expansion fit of the reduced-ITD introduced by 10\% variation in $\alpha_s$ at $\mu=2.0$ GeV.
\begin{table*}[!htp]
  \centering
  \begin{tabular}{|c|c|c|c|c|c|c|c|}
  \hline
    $\lambda_1$  & $\lambda_2$  & $\lambda_3$ &  $\lambda_4$ & $c_1$ & $c_2$ & $c_3$ & $\chi^2/{\rm d.o.f}$\\
    \hhline{|=|=|=|=|=|=|=|=|}
    $-0.0083(49)$ & $-0.79(16)$ & -2.87(1.35) & -7.19(2.87) & -0.00080(97) & -0.00014(16) & -0.11(19) & 0.72 \\
   \hhline{|=|=|=|=|=|=|=|=|}
  \end{tabular}
  \caption{The parameters of  the correlated simultaneous fit to obtain $Q(\nu)$ from the two ensembles. }
  \label{tab:zfit}
\end{table*}

We now use the $Q(\nu,\mu=2$ GeV) ITD  from the above fit to extract  the pion valence quark distribution. By definition, $Q(\nu)$ and the valence quark distribution of the pion $q^\pi_{\rm v}(x)$ are related by
\bea
Q(\nu) = \int_{-1}^1 dx \,q_{\rm v}(x)\, e^{i\nu x}\ , 
\eea
and the quark distribution is given by the inverse Fourier transform
\bea \la{inverseFT}
q_{\rm v}(x) = \frac{1}{2\pi} \int_{-\infty}^{\infty} d\nu e^{-i\nu x} Q(\nu).
\eea
Therefore to convert $Q(\nu)$ into a function of $x$, one should, in principle, know $Q(\nu)$ for all $\nu$. In our lattice QCD calculation, we are
    restricted to $\nu_{\rm max}\sim 6$ for a discrete set of integer
    $\nu$.  Thus, the extraction of the PDF using Eq.~\eqref{inverseFT} from lattice calculated data constitutes an ill-posed inverse problem. To our knowledge, a reliable direct inverse Fourier transform to extract the PDF using  lattice QCD data is  currently a formidable task.

An important constraint serving as additional information 
is  that the valence distributions of  the nucleon and  the pion are smooth functions of the momentum fraction $x$ in the region $0<x<1$, with support only in that region, and the pseudo-ITD is related  by a cosine transform  to the pion valence quark distribution~\cite{Braun:1994jq,Orginos:2017kos}
\bea \la{costran}
Q(\nu) = \int_0^1 dx \cos(\nu x)\, q_{\rm v}^\pi(x).
\eea
In the spirit of the functional forms used in global fits of PDFs, we insert
\bea \label{pdf-form}
q^\pi_{\rm v}(x) = N x^\alpha(1-x)^\beta(1+ \rho \sqrt{x} + \gamma x)
\eea
into Eq.~\eqref{costran} and numerically perform the integration, where $N$ is the normalization such that
\bea \label{condition}
\int_0^1 dx\,q^\pi_{\rm v}(x) =1.
\eea

 \begin{figure}[!htp]
  \centering
  \subfigure[]{\includegraphics[width=0.48\textwidth]{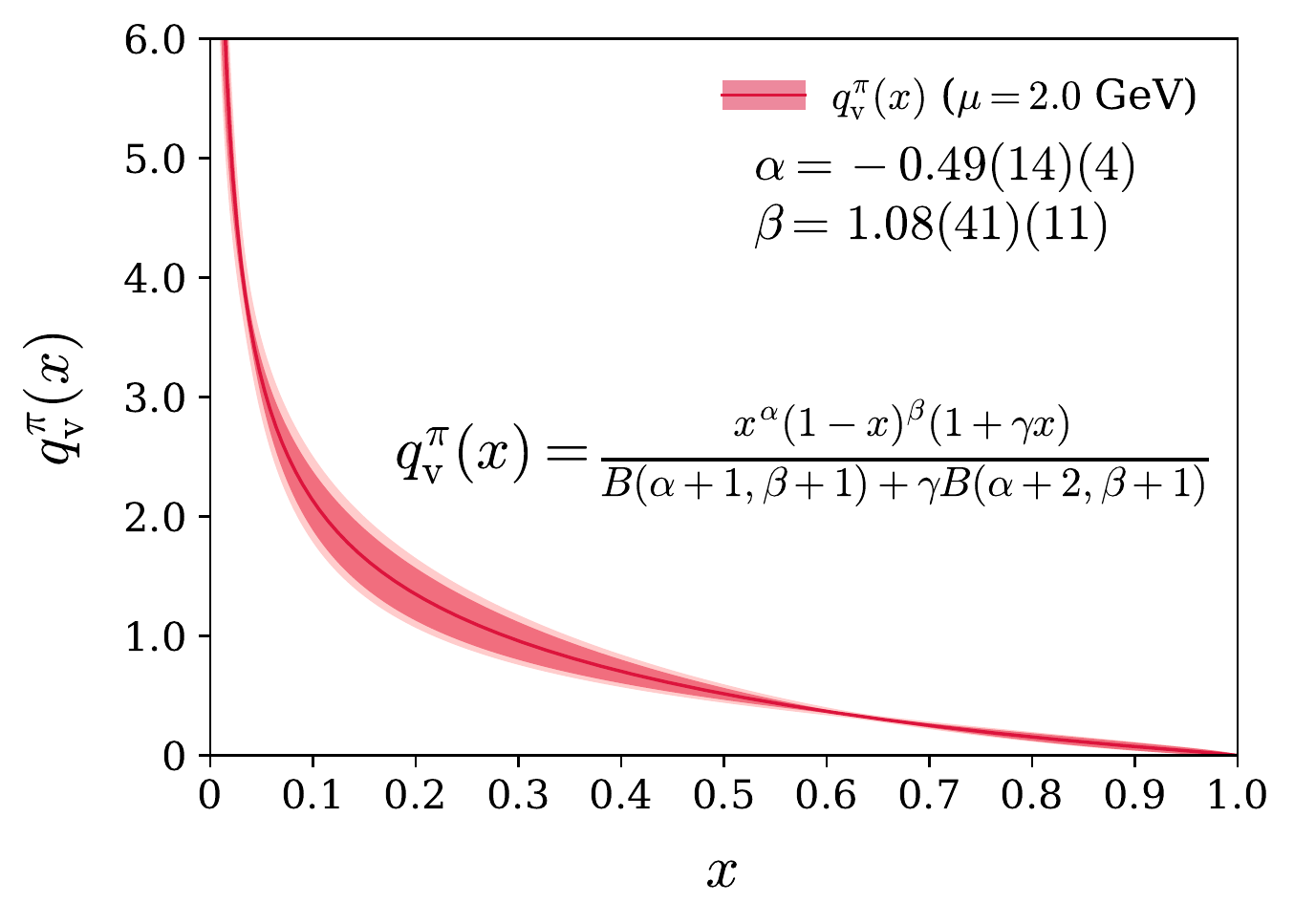}\label{4a}}
  \subfigure[]{\includegraphics[width=0.48\textwidth]{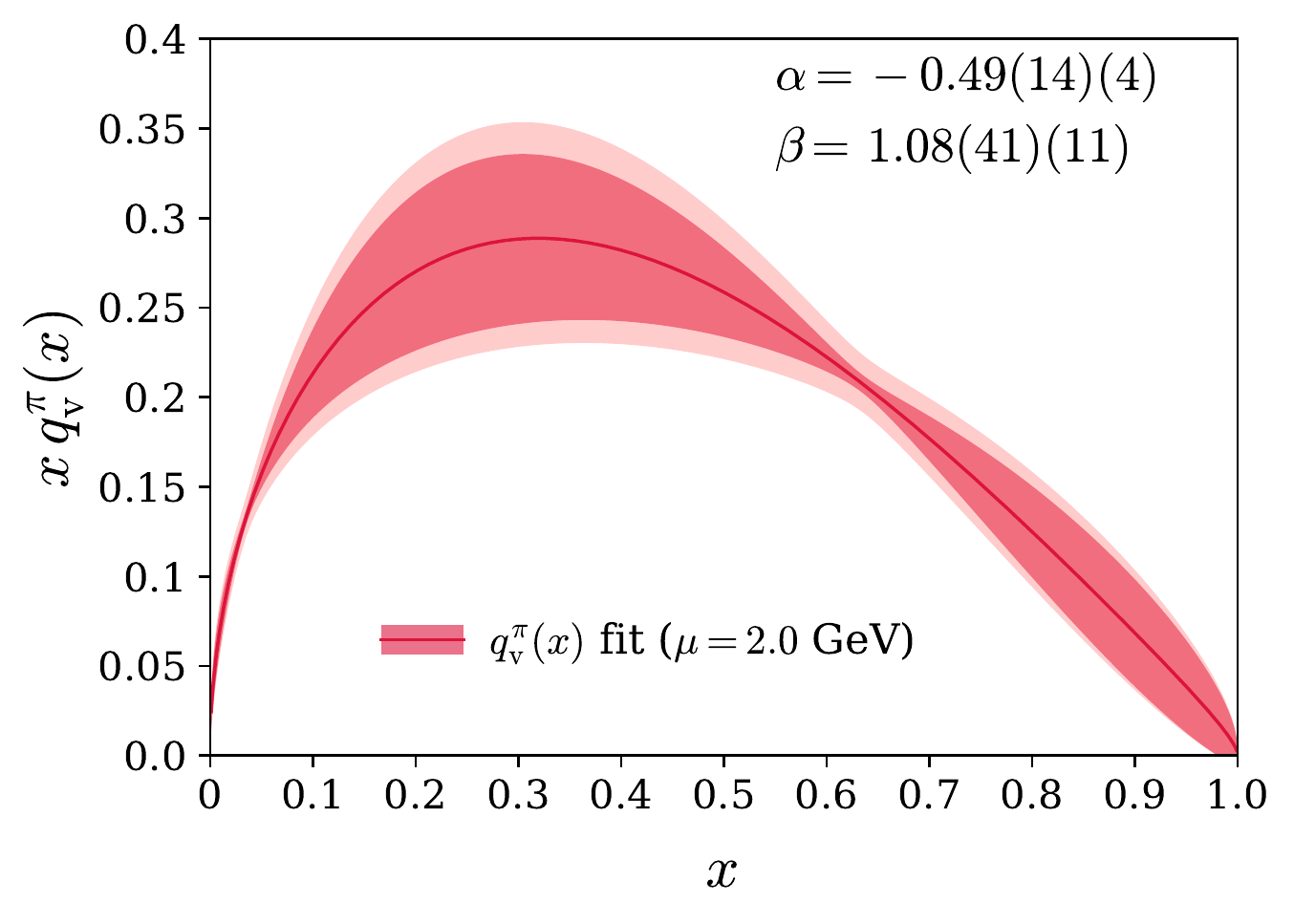}\label{4b}}
  \caption{\label{PDFres}
  The pion valence distribution obtained from the fit in Eq.~\eqref{pdf-form-use} using the NLO perturbative kernel in Eq.~\eqref{matching}. FIG.~\ref{4a} shows the pion valence distribution $q^\pi_{\rm v}(x)$ and FIG.~\ref{4b}  shows the $xq^\pi_{\rm v}(x)$-distribution.  The initial scale for evolving the PDF to a higher scale is  $\mu=2$ GeV, as described in  Section~\ref{comparison}. For the fits of PDFs, we use the covariance matrix obtained from the $z$-expansion fit to generate bootstrap samples in the Ioffe-time range of $0< \nu <4.71$ for which lattice QCD data points exist. We do not perform any extrapolation to the data points outside this region of the Ioffe-time using the results of the $z$-expansion. We have checked that there is no dependence on the number of bootstrap samples in our fit. The statistical uncertainty band is obtained from the fits  to  the bootstrap samples of the data. The inner red band is obtained from a simultaneous fit to the matched ITDs at $\mu=2.0$ GeV on these two ensembles in the limit of infinite volume. The outer red uncertainty band in the extraction of PDFs is obtained as a source of systematic uncertainty by calculating the difference between the simultaneous fit to the matched ITDs  for $\alpha_s=0.303\pm 0.030$. }
\end{figure}

\begin{table}[!htb]
  \centering
  \begin{tabular}{| c | c | c | }
  \hline
 n & $\langle x^n  \rangle$ (This calculation) & $\langle x^n  \rangle$ WRH~\cite{Wijesooriya:2005ir} \\
    \hhline{|=|=|=|}
    1 & 0.165(9) & 0.217(11) \\
    2 & 0.064(1) & 0.087(5) \\
    3 & 0.033(2)  & 0.043(3) \\
    4 & 0.020(2)  & $-$ \\
    \hhline{|=|=|=|}
  \end{tabular}
  \caption{Comparison between the moments of the PDF extracted from the pion valence distribution in this work and those obtained in the NLO QCD analysis~\cite{Wijesooriya:2005ir} using the Fermilab experimental data~\cite{Conway:1989fs} at a scale of 5.2 GeV. }
  \label{tab:mom27}
\end{table}

We use the numerical fitting program, ROOT~\cite{ROOT} to fit bootstrap  samples of the matched pseudo-ITD to  the $q^\pi_{\rm v}(x)$-distribution. We find that  the $\rho \sqrt{x}$ or the $\gamma x$ term has no effect in the fit and we do not obtain any signal of the fit parameters $\rho$ and $\gamma$. Therefore we drop these terms and adopt in our calculations the following simple functional form for the PDF,
\bea \label{pdf-form-use}
q^\pi_{\rm v}(x) = \frac{x^\alpha(1-x)^\beta}{B(\alpha+1,\beta+1)},\nn \\
\eea
where the beta functions in the denominator ensure that the normalization condition in Eq.~\eqref{condition} is met.  We obtain the fit parameters,
\bea \la{params}
\alpha &=& -0.48(14)_{\rm stat}(4)_{\rm sys}, \nn \\
\beta & = & 1.08(41)_{\rm stat}(11)_{\rm sys},
\eea 
with the $\chi^2/{\rm d.o.f.}$ about 1.9.  For example, the inclusion of the $\gamma x$ term in the PDF fit yields the following fit parameters and therefore the extracted PDF remains essentially unchanged.
\bea \la{params2}
\alpha &=& -0.49(14)_{\rm stat}(2)_{\rm sys}, \nn \\
\beta & = & 1.05(37)_{\rm stat}(2)_{\rm sys}, \nn \\
\gamma &=& 0.003(11)_{\rm stat}(57)_{\rm sys}.
\eea 

In Eqs.~\eqref{params} and~\eqref{params2}, the number in the first uncertainty in the parentheses of the fit parameters are statistical uncertainties and the second is obtained from fitting the matched Ioffe time distribution with the 10\% variation in the the value of $\alpha_s$.  We present the extracted PDF $q^\pi_{\rm v}(x)$ from this fit in FIG.~\ref{4a} and the $xq^\pi_{\rm v}(x)$-distribution in FIG.~\ref{4b}. The fit to the data returns a
  well-constrained value, $\alpha = -0.48(14)$, for the small-$x$
  behavior of the PDF corresponding to the slope of the relevant Regge
  trajectory.  However, we stress that, with the present resources of
  lattice QCD calculations, a precise and accurate determination of
  the low-$x$ behavior of the PDFs is not accessible. Specifically, one can argue that the Fourier transform at the Ioffe time $\nu$ is related to the region around the inverse of
  the Bjorken variable $x_{\rm B}$, {\it i.e.} $\nu=1/x_{\rm B}$~\cite{Ma:2017pxb}. Therefore, to obtain a reliable estimate of
  the low-$x$ behavior of {the} PDFs, one requires knowledge of the
  ITD at large $\nu$.  The PDF model parameters of the fit are highly correlated, as is evident from FIGs.~\ref{4a} and~\ref{4b} . The uncertainty at $x=0.65$ shrinks significantly due to the correlation between the fit parameters. This feature of shrinking uncertainty at different $x$-values has also been observed in the calculation of nucleon PDFs using pseudo-PDFs approach~\cite{Joo:2019jct}. It is a feature of these highly correlated fits to have regions with small statistical errors.

For a comparison with the ``OPE without OPE" calculation of the moments in Table~\ref{tab:mom}, we can take the Mellin transformation of this PDF result described by the fit parameters in Eq.~\eqref{params}. At the scale of $\mu=2$ GeV, the first four moments are $\langle x\rangle = 0.188(56)$, $\langle x^2\rangle = 0.081(29)$, $\langle x^3\rangle = 0.046(19)$, and $\langle x^4\rangle = 0.030(14)$. Note that, any discrepancy between these numbers and those in Table~\ref{tab:mom} may arise from a number of different reasons. Firstly, the odd moments quoted in Table~\ref{tab:mom} contain small contribution from the antiquarks. Secondly, the data set used in the PDF extraction includes data from both ensembles while those in Table~\ref{tab:mom} only include data from the $a127m415L$ ensemble. We also present a comparison between the moments extracted from our pion valence PDF fit with those extracted in the NLO QCD analysis~\cite{Wijesooriya:2005ir} using the Fermilab experimental data~\cite{Conway:1989fs} at a scale of 5.2 GeV in the Table~\ref{tab:mom27}. We note that, this lattice QCD calculation is performed at an unphysical pion mass of $m_\pi\sim 415$ MeV. A proper comparison with the QCD analysis of the experimental data and lattice QCD calculation can be made when a continuum and infinite volume extrapolation to the lattice data near the physical pion mass is performed.


\section{Comparison with Other Determinations} \la{comparison}

This lattice QCD calculation using the pseudo-ITD approach is performed at a relatively heavy pion mass ($m_\pi\simeq415\text{ MeV}$) and on a relatively coarse lattice spacing of $a=0.127$ fm.  Repeating similar calculations on several other lattice ensembles to determine the pion mass dependence, quantify the severity of discretization errors and provide a more precise estimation of the effect of finite volume in order to obtain the pion valence PDF in the continuum limit is under way. However, with the present calculation, we proceed with a qualitative comparison of  various global fits of the pion PDFs and three previous lattice QCD determinations.

For a comparison with $q^\pi_{\rm v}(x)$ determined from the Drell-Yan experimental data in Ref.~\cite{Conway:1989fs}, we evolve our determination of the pion PDF to an evolution scale of \mbox{$\mu^2=27$ GeV$^2$} starting from an initial scale of \mbox{$\mu_0^2=4$ GeV$^2$} as shown in FIGs.~\ref{4a} and~\ref{4b}. As expected, the evolution to a higher scale shifts the peak of the $xq^\pi_{\rm v}(x)$-distribution toward smaller values of $x$ and a more convex-up behavior of the distribution as $x\to1$ is seen compared to the $xq^\pi_{\rm v}(x)$ at the initial scale of our calculation.  

\begin{figure}[htp]
\begin{center}
\includegraphics[width=3.5in]{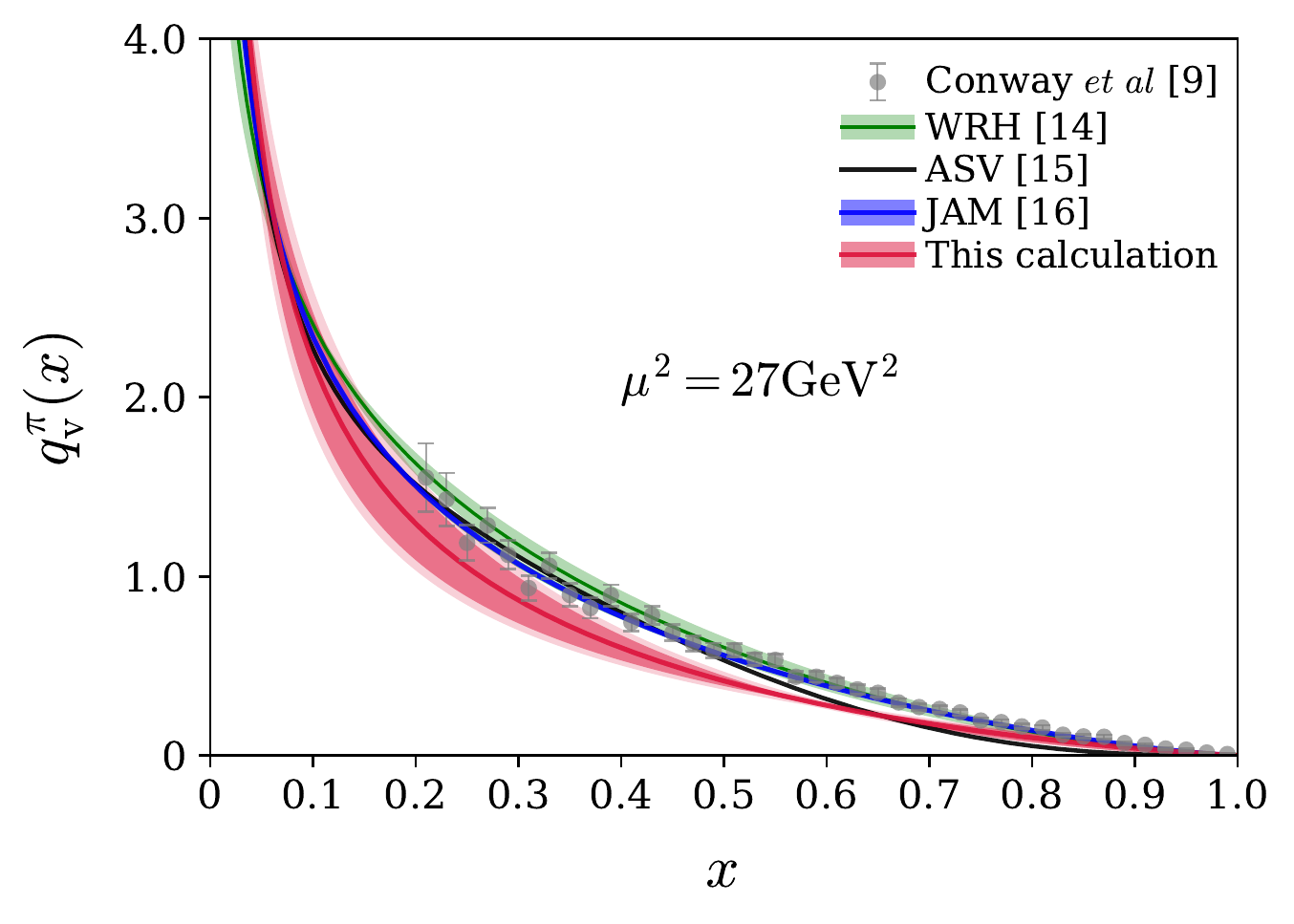}
\caption{\label{fig:evo-noX}
Comparison of the pion $q^\pi_{\rm v}(x)$-distribution with the leading-order (LO) extraction from Drell-Yan data~\cite{Conway:1989fs} (gray data points with uncertainties), next-to-leading order (NLO) fits~\cite{Wijesooriya:2005ir,Aicher:2010cb,Barry:2018ort} (green band, maroon curve, and blue band). This lattice QCD calculation of $q^\pi_{\rm v}(x)$ is evolved from an initial scale $\mu^2=4$ GeV$^2$ at NLO. All the results are at evolved to an evolution scale of $\mu^2=27$ GeV$^2$. The outer red uncertainty band shown in the $q^\pi_{\rm v}(x)$-distribution is obtained from the variation in the choice of $\alpha_s$ during the one-loop perturbative matching as described in Section~\ref{results}. }
\end{center}
\end{figure}

\begin{figure}[htp]
\begin{center}
\includegraphics[width=3.5in]{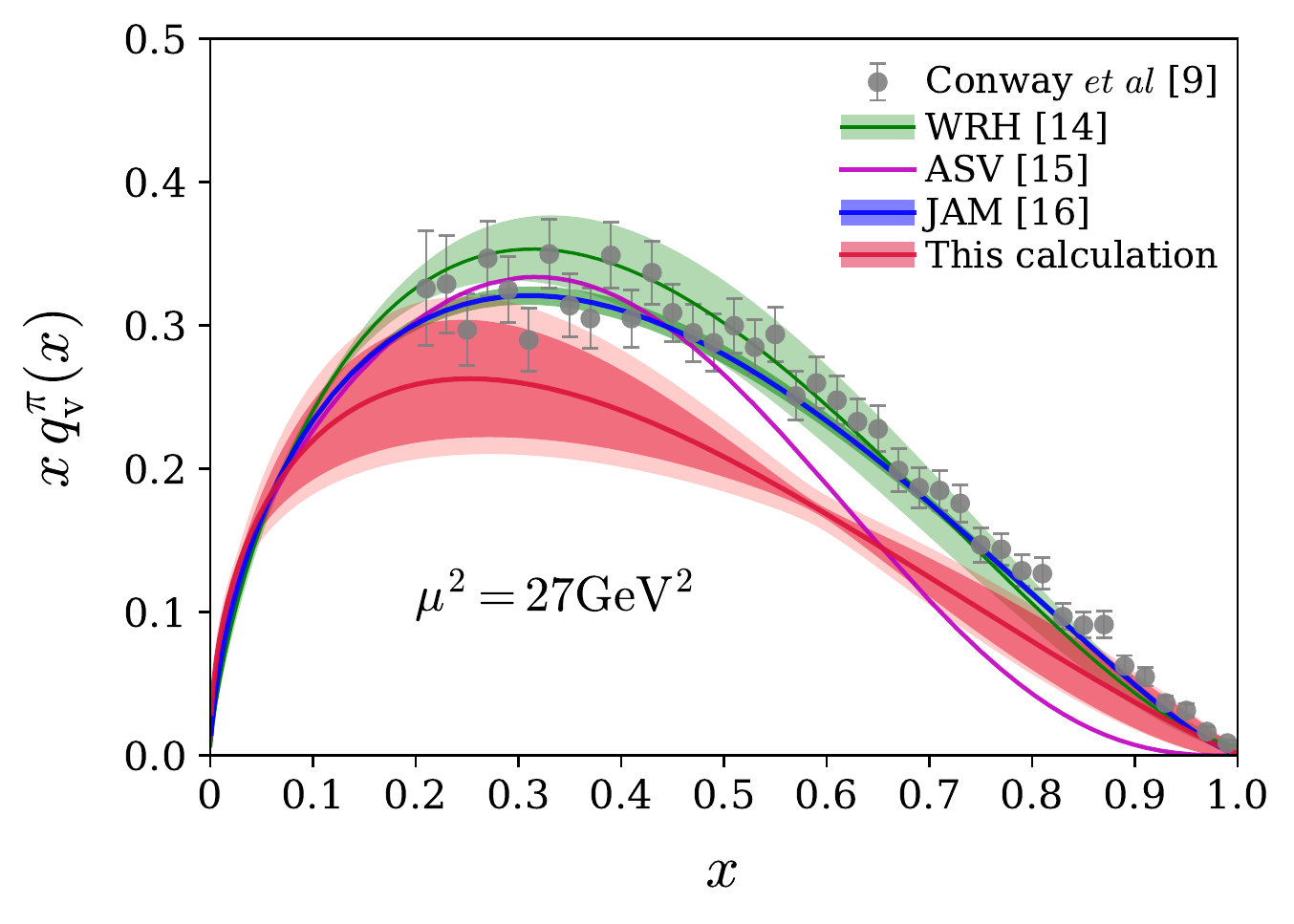}
\caption{\label{fig:evolPDF}
Comparison of the pion $xq^\pi_{\rm v}(x)$-distribution with the LO extraction from Drell-Yan data~\cite{Conway:1989fs} (gray data points with uncertainties), NLO 
 fits~\cite{Wijesooriya:2005ir,Aicher:2010cb,Barry:2018ort} (green band, maroon curve, and blue band). This lattice QCD calculation of $q^\pi_{\rm v}(x)$ is evolved from an initial scale $\mu_0^2=4$ GeV$^2$ at NLO. All the results are  evolved to an evolution scale of $\mu^2=27$ GeV$^2$. The outer red band shown in the $q^\pi_{\rm v}(x)$-distribution is obtained from the variation in the choice of $\alpha_s$ during the one-loop perturbative matching as described in Section~\ref{results} and by calculating the variation in the fitting of the matched Ioffe-time data using the PDFs parametrization in Eq.~\eqref{pdf-form-use}. }
\end{center}
\end{figure}

We see a quantitative agreement of our extracted pion PDF with the NLO global fits in~\cite{Wijesooriya:2005ir,Barry:2018ort} to the Drell-Yan experimental data in~\cite{Conway:1989fs} in the $x\gtrsim 0.7$ momentum fraction region as presented in FIG.~\ref{fig:evo-noX}.  It should be noted that our determination of $q^\pi_{\rm v}(x)$ has a distinctive deviation from the predictions of QCD-based hard-gluon-exchange perturbative models~\cite{Farrar:1979aw,Berger:1979du,Brodsky:1994kg}, which predicts a faster $(1-x)^2$ fall-off of the pion PDF at large $x$.  Such a  $(1-x)^2$ fall-off at large $x$ was also obtained in the experimental data analysis in Ref.~\cite{Aicher:2010cb}, where the authors included next-to-leading-logarithmic threshold soft-gluon re-summation effects in the  calculation of the Drell-Yan cross section. 
A good way to visualize  the discrepancy in the large-$x$ region between the pion PDF extracted in our calculation with the experimental data and various global fits can be demonstrated by plotting $xq^\pi_{\rm v}(x)$ as a function of $x$. We present such a plot in FIG.~\ref{fig:evolPDF}.

We now compare the determination of the pion valence quark PDF presented in this paper with the previous three lattice QCD determinations using the quasi-PDF approach~\cite{Chen:2018fwa,Izubuchi:2019lyk} and the lattice cross sections approach~\cite{Sufian:2019bol}. 

In~\cite{Izubuchi:2019lyk}, a careful and systematic investigation was performed. In that paper,
the  matrix element was renormalized using the  RI/MOM inspired 
approach~\cite{Alexandrou:2017huk, Chen:2017mzz}, in 
which the hadron matrix element is  divided  by a quark matrix element of the same operator.
Thus, just like in the reduced pseudo-ITD method, one deals with a ratio
of the original matrix element and another matrix element that 
has the same ultraviolet divergences. 

However, the nonperturbative
$z^2$-behavior of the denominator factor in these two approaches is different.
In particular, according to the analysis in Ref.~\cite{Izubuchi:2019lyk},
the RI/MOM  factor $Z_{\rm RI/MOM} (z)$ in the $z\gtrsim0.5$ fm region shows a formation 
of a constituent quark mass \mbox{$m_{\rm scr} \sim 300$ MeV}
leading to a  suppression of $Z_{\rm RI/MOM}(z)$ by an extra $e^{-m_{\rm scr}  |z|}$ factor.

 The pion rest-frame matrix element ${\mathcal M} (0, z^2)$  used in our calculation
 has a much faster decrease with $|z|$ than $Z_{\rm RI/MOM}(z)$ of Ref.~\cite{Izubuchi:2019lyk}.
 Numerically, it is very close to the nucleon rest-frame matrix element
 of Ref.~\cite{Joo:2019jct}.  As argued in Ref.~\cite{Radyushkin:2017cyf},
 the fast fall-off of ${\mathcal M} (0, z^2)$ reflects the finite size of the relevant hadron,
 i.e. the nonperturbative effects related to quark confinement. 
 In the OPE language, the associated $z^2$-dependence corresponds 
 to higher-twist contributions. 
 As pointed out in Section~\ref{method}, one of the aims of 
 using  the reduced ITD is to   cancel unwanted higher twist effects.
 
 We note  that the use of the  reduced ITD corresponds 
 to a gauge-invariant renormalization prescription that 
 avoids  the  pathological systematics of  fixed gauge renormalization.
In fact, previous calculations of quasi-PDFs~\cite{Chen:2018fwa,Alexandrou:2019lfo} have shown slight discrepancies which depend on the renormalization scheme used for the matrix element and intermediate schemes used in the matching relationships. The systematic errors introduced by these choices can be avoided by calculating a renormalization group invariant quantity 
${\mathfrak M} (\nu, z^2)$ and then matching it to the $\overline{\mbox{MS}}$ PDF. 
 
Another difference among the different lattice calculations is the treatment of the inverse problem. 
Our point is that inverse problems can only be solved by adding additional information. The quasi-PDF calculation in~\cite{Izubuchi:2019lyk} adds this information in a  way
analogous to the present  calculation and the lattice cross sections
 calculation in~\cite{Sufian:2019bol}. The main idea is to  parameterize the 
 PDF  in terms of a few model dependent parameters and then fit the  position-space 
 matrix element  using that functional form. 
 
 The quasi-PDF calculation performed in Ref.~\cite{Chen:2018fwa} instead attempts to directly perform the inverse Fourier transform. To remove unphysical oscillations caused by the ill-posed inverse, they use a ``derivative method''~\cite{Lin:2017ani}. As was shown in~\cite{Karpie:2018zaz}, the ``derivative method'' does not alleviate the ill-posed inverse problem especially with such short Ioffe time ranges, as only a few nonzero points exist in that quasi-PDF calculation~\cite{Chen:2018fwa}. 
 
 Of notable interest, our calculation of $x q^\pi_{\rm v}(x)$, shown in FIGs.~\ref{PDFres} and~\ref{fig:evolPDF}, illustrates a peak of the distribution in a region $x<0.40$. This is consistent with all the global analyses of the pion valence distribution, wherein $x q^\pi_{\rm v}(x)$ is peaked below $x=0.40$. This feature also occurs in the lattice cross sections calculation in~\cite{Sufian:2019bol} and the quasi-PDF calculation in~\cite{Izubuchi:2019lyk}. On the other hand, the quasi-PDF calculation using the derivative method performed in~\cite{Chen:2018fwa} peaks at a somewhat larger value of around \mbox{$x=0.50$}.

In comparison with the determination of $q_{\rm v}^\pi(x)$  in Ref.~\cite{Sufian:2019bol}, which was performed on the same ensemble as the larger of our lattices, we see the large-$x$ behavior of the distributions are in agreement within their uncertainties. In particular,
$q_{\rm v}^\pi(x)$ extracted in~\cite{Sufian:2019bol}  gives 
 $\beta =  1.93(68)$, and 
 the value   $\beta = 1.08(41)$ is obtained in the present calculation. We note, however,  that the choice of 
 the initial scale of  \mbox{1 GeV} performed in Ref.~\cite{Sufian:2019bol}   was
 rather  arbitrary because of the absence of the NLO perturbative matching. The difference in $q_{\rm v}^\pi(x)$ obtained in these two calculations based on short-distance factorization remains to be investgated. 

\begin{figure}[t]
  \subfigure[]{\includegraphics[width=0.50\textwidth]{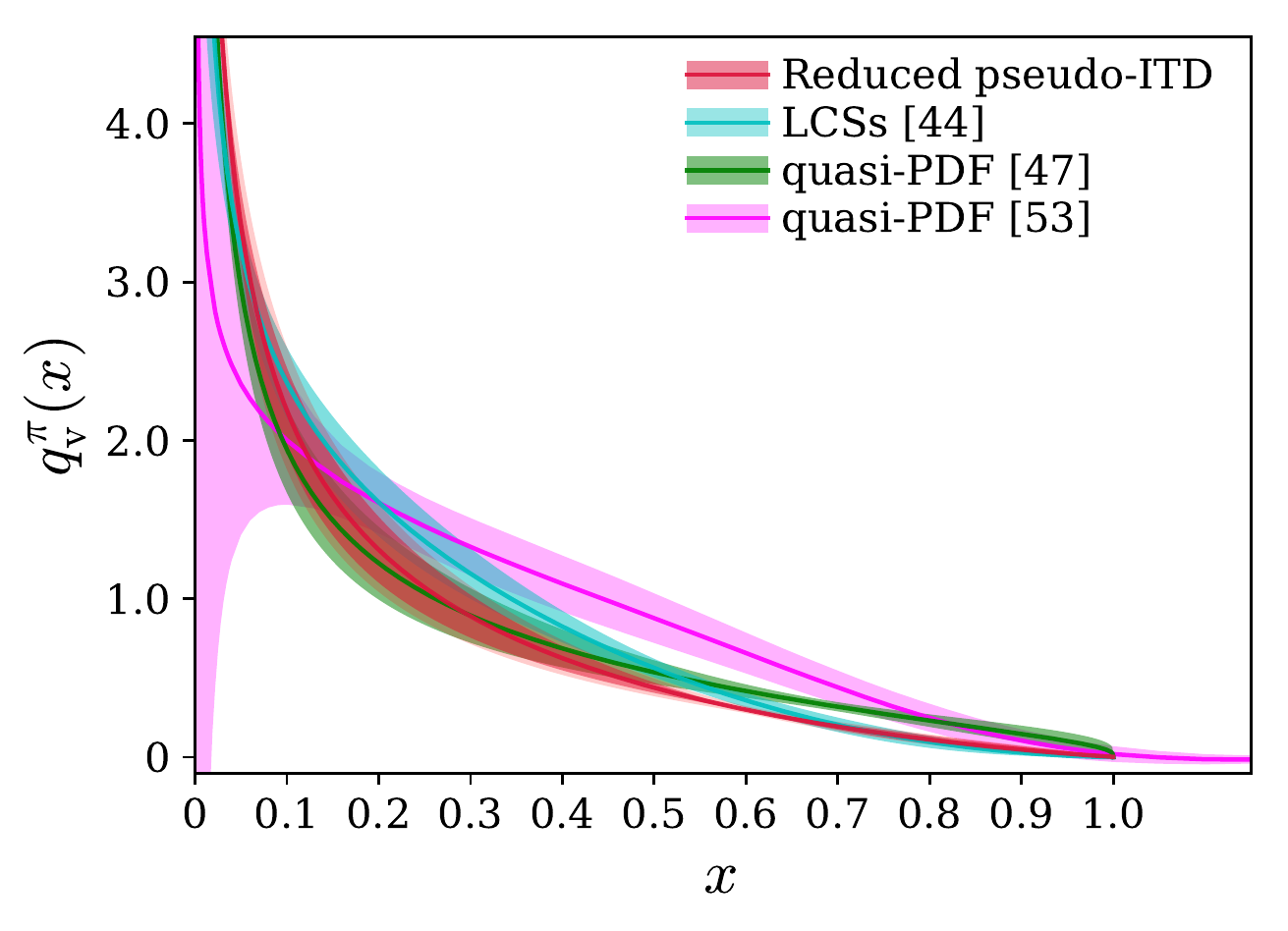}\label{9a}}
  \subfigure[]{\includegraphics[width=0.50\textwidth]{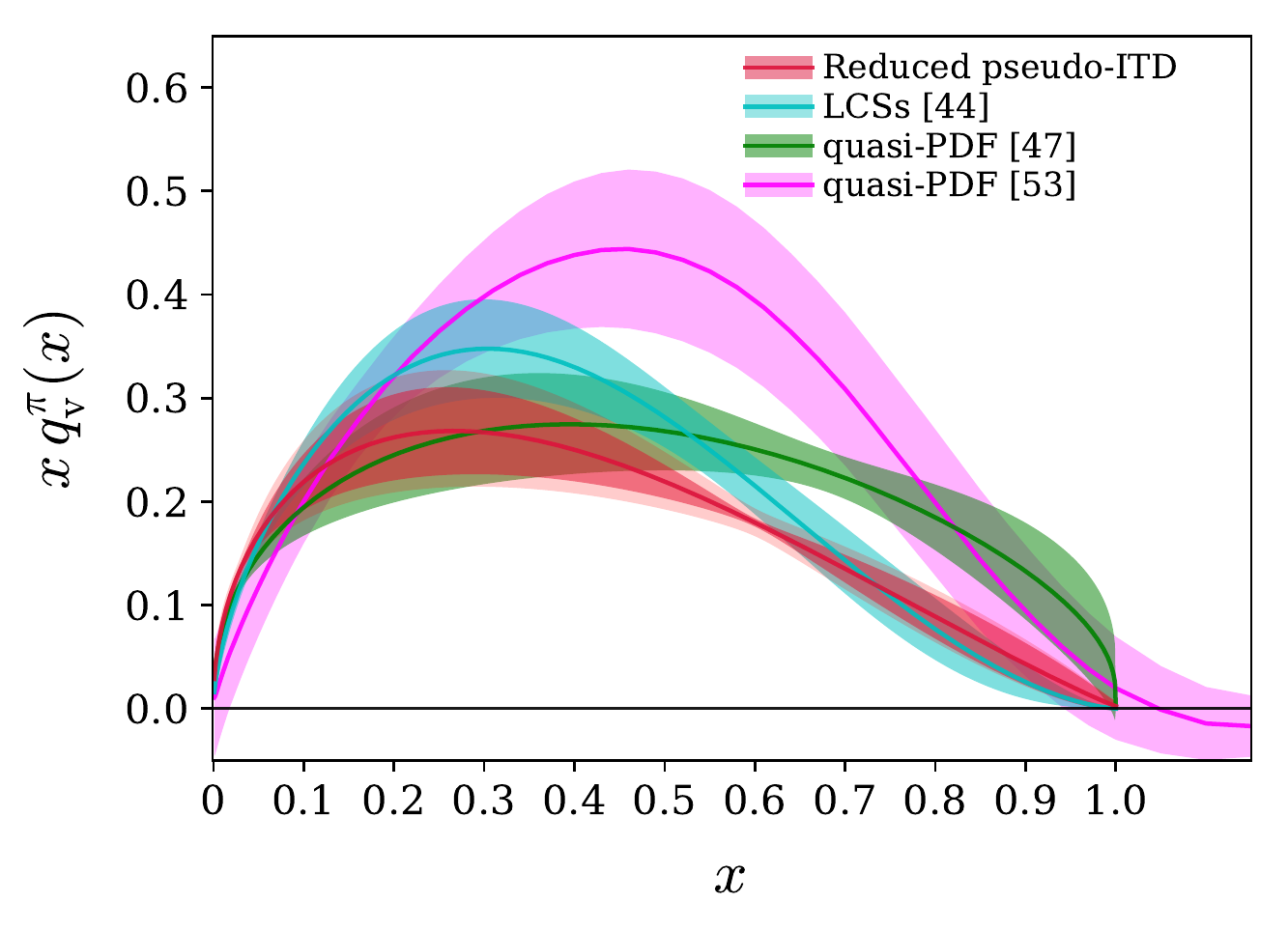}\label{9b}}
  \caption{\label{fig:latcomparison}
   Comparison of the pion valence $q^\pi_{\rm v}(x)$ and $xq^\pi_{\rm v}(x)$-distributions estimated near $\mu=4$ GeV scale using quasi-PDFs in Refs.~\cite{Chen:2018fwa,Izubuchi:2019lyk}, good lattice cross sections in Ref.~\cite{Sufian:2019bol} with the pseudo-ITD approach in this calculation.} 
\end{figure}

 In FIG.~\ref{fig:latcomparison}, we present a comparison between the lattice QCD extractions of pion valence quark distribution made in  Refs.~\cite{Chen:2018fwa,Izubuchi:2019lyk,Sufian:2019bol} with our  calculation that uses the pseudo-ITD approach. For the calculation in Ref.~\cite{Sufian:2019bol}, the PDF is evolved to $\mu=4$ GeV assuming an initial scale of 2 GeV in FIG.~\ref{fig:latcomparison}. The PDF in Ref.~\cite{Izubuchi:2019lyk} is calculated using pion momentum $p_z=1.29$ GeV,
 the RI/MOM scale is fixed at 1.93 GeV and the PDF is estimated at $\mu=3.2$ GeV. To illustrate  the difference between all the  calculations in the $x\gtrsim 0.4$ region, we also present the $xq^\pi_{\rm v}(x)$ distributions from these lattice calculations in FIG.~\ref{9b}.


\section{Summary and Outlook}

In this paper, we have presented the first lattice QCD calculation of the pion valence distribution  using the Ioffe time pseudo distributions approach. In our calculation, we have used combined  results from two different ensembles,
with the same lattice spacing \mbox{$a=0.127$ fm,} but with different lattice sizes  of 
$24^3\times 64$ and $32^3\times 96$. We have also performed the first lattice QCD calculation of the fourth moment of the pion PDF, which was previously inaccessible from local matrix elements.

We have performed the  one-loop perturbative matching 
of the pseudo-ITDs to light-cone ITDs at the scale $\mu =2$ GeV.  Then we have combined  the pseudo-distributions from the two ensembles using a   model-independent scheme. 

To   extract the pion valence quark distribution, 
we have assumed a  functional form motivated by those  used  in phenomenological global fits of parton distribution functions,
and obtained the parameters of this form using our data for the matched light-cone  ITD. 
This approach allows us to avoid solving the ill-posed problem of the  inverse Fourier transform
from the ITD to PDF. 
We made  a qualitative comparison between our lattice QCD extraction of the pion valence quark distribution with those obtained from global fits and previous lattice QCD calculations.  

It should be noted that, in our lattice calculation,  the  region $z\lesssim 0.5$ fm where one can rely on the perturbative matching corresponds just to 4 lattice separations at this lattice spacing. Therefore, a calculation with finer lattice spacing is 
warranted and is the goal of future efforts. 
On the other hand, our analysis  (even keeping in mind the limitations of the present statistics) shows no 
 evidence for higher twist effects in the reduced pseudo-ITD   up to $z=1$ fm, which we took as a 
 justification to  use  the data for extracting the PDF up to $z=8a$ on the larger ensemble.
 
  A natural extension of  our   investigation  would be  a calculation on several other ensembles with different pion masses, lattice spacings, and volumes.  We expect  that the results 
  of such upgraded  lattice QCD  calculations  will  eventually become  important additional input  in global analyses to obtain a precise knowledge of the large-$x$ behavior of the pion valence quark distribution.

\section*{Acknowledgments}

R.S.S. thanks Nikhil Karthik, Luka Leskovec, Tianbo Liu, Charles Shugert, and Jian-Hui Zhang for useful discussions. 
 This work is supported by Jefferson
Science Associates, LLC under U.S. DOE Contract \#DE-AC05-06OR23177 within the framework of the TMD Collaboration.  
We acknowledge the facilities of the USQCD Collaboration
used for this research in part, which are funded by the Office of Science of the U.S. Department
of Energy. This material is based in part upon work supported by a grant from the Southeastern Universities Research Association (SURA) under an appropriation from the Commonwealth of Virginia. This work used the Extreme Science and Engineering Discovery Environment (XSEDE), which is supported by National Science Foundation grant number ACI-1548562~\cite{xsede}. This work was performed in part using computing facilities at the College of William and Mary which were provided by contributions from the National Science Foundation (MRI grant PHY-1626177), the Commonwealth of Virginia Equipment Trust Fund and the Office of Naval Research. In addition, this work used resources at NERSC, a DOE Office of Science User Facility supported by the Office of Science of the U.S. Department of Energy under Contract \#DE-AC02-05CH11231, as well as resources of the Oak Ridge Leadership Computing Facility at the Oak Ridge National Laboratory, which is supported by the Office of Science of the U.S. Department of Energy under Contract No. \#DE-AC05-00OR22725. The authors gratefully acknowledge the computing time
granted by the John von Neumann Institute for Computing (NIC) and
provided on the supercomputer JURECA at J\"{u}lich Supercomputing Centre
(JSC)~\cite{jureca}.  J.K. is supported in part by the U.S. Department of Energy under contract DE-FG02-04ER41302 and Department of Energy Office of Science Graduate Student Research fellowships, through the U.S. Department of Energy, Office of Science, Office of Workforce Development for Teachers and Scientists, Office of Science Graduate Student Research (SCGSR) program. K.O. acknowledges support in part  by the U.S. Department of Energy through Grant Number DE- FG02-04ER41302, by STFC consolidated grant ST/P000681/1. 
 A.R. was supported in part by U.S. DOE Grant
\mbox{\#DE-FG02-97ER41028}. SZ acknowledges support by the DFG
Collaborative Research Centre SFB 1225 (ISOQUANT).

\providecommand{\href}[2]{#2}
\begingroup\raggedright

\endgroup

\end{document}